\documentclass[conference]{IEEEtran}
\IEEEoverridecommandlockouts
\usepackage{algorithmic}
\usepackage{graphicx}
\usepackage{textcomp}
\usepackage{xcolor}
\usepackage{booktabs}
\usepackage[hidelinks]{hyperref}

\usepackage{subcaption}
\usepackage{xcolor}
\usepackage{subfiles}
\usepackage[utf8]{inputenc}

\usepackage{breakurl} 
\usepackage{cite}

\usepackage{url, xspace,relsize}
\hypersetup{
    colorlinks = true,
    linkcolor = blue,
    anchorcolor = blue,
    citecolor = blue,
    filecolor = blue,
    urlcolor = blue}

\usepackage{listings}
\definecolor{verylightgray}{rgb}{.97,.97,.97}
\lstdefinelanguage{Solidity}{
	keywords=[1]{anonymous, assembly, assert, balance, break, call, callcode, case, catch, class, constant, continue, constructor, contract, debugger, default, delegatecall, delete, do, else, emit, event, experimental, export, external, false, finally, for, function, gas, if, implements, import, in, indexed, instanceof, interface, internal, is, length, library, log0, log1, log2, log3, log4, memory, modifier, new, payable, pragma, private, protected, public, pure, push, require, return, returns, revert, selfdestruct, send, solidity, storage, struct, suicide, super, switch, then, this, throw, transfer, true, try, typeof, using, value, view, while, with, addmod, ecrecover, keccak256, mulmod, ripemd160, sha256, sha3}, 
	keywordstyle=[1]\color{blue}\bfseries,
	keywords=[2]{address, bool, byte, bytes, bytes1, bytes2, bytes3, bytes4, bytes5, bytes6, bytes7, bytes8, bytes9, bytes10, bytes11, bytes12, bytes13, bytes14, bytes15, bytes16, bytes17, bytes18, bytes19, bytes20, bytes21, bytes22, bytes23, bytes24, bytes25, bytes26, bytes27, bytes28, bytes29, bytes30, bytes31, bytes32, enum, int, int8, int16, int24, int32, int40, int48, int56, int64, int72, int80, int88, int96, int104, int112, int120, int128, int136, int144, int152, int160, int168, int176, int184, int192, int200, int208, int216, int224, int232, int240, int248, int256, mapping, string, uint, uint8, uint16, uint24, uint32, uint40, uint48, uint56, uint64, uint72, uint80, uint88, uint96, uint104, uint112, uint120, uint128, uint136, uint144, uint152, uint160, uint168, uint176, uint184, uint192, uint200, uint208, uint216, uint224, uint232, uint240, uint248, uint256, var, void, ether, finney, szabo, wei, days, hours, minutes, seconds, weeks, years},	
	keywordstyle=[2]\color{teal}\bfseries,
	keywords=[3]{block, blockhash, coinbase, difficulty, gaslimit, number, timestamp, msg, data, gas, sender, sig, value, now, tx, gasprice, origin},	
	keywordstyle=[3]\color{violet}\bfseries,
	identifierstyle=\color{black},
	sensitive=false,
	comment=[l]{//},
	morecomment=[s]{/*}{*/},
	commentstyle=\color{gray}\ttfamily,
	stringstyle=\color{red}\ttfamily,
	morestring=[b]',
	morestring=[b]"
}

\usepackage{varwidth}
\usepackage{pifont}

\newcommand{\bheading}[1]{{\vspace{3pt}\noindent{\textbf{#1}}}}
\newcommand{\iheading}[1]{{\vspace{2pt}\noindent{\textit{#1}}}}

\newcommand{\cc}[1]{\mbox{\smaller[0.5]\texttt{#1}}}

\newenvironment{packeditemize}{
\begin{list}{$\bullet$}{
\setlength{\labelwidth}{8pt}
\setlength{\itemsep}{0pt}
\setlength{\leftmargin}{\labelwidth}
\addtolength{\leftmargin}{\labelsep}
\setlength{\parindent}{0pt}
\setlength{\listparindent}{\parindent}
\setlength{\parsep}{0pt}
\setlength{\topsep}{3pt}}}{\end{list}}

\usepackage{xspace}
\newcommand{\etal}{\emph{et al.}\xspace}

\newcommand{\ie}{\emph{i.e.}\xspace}
\newcommand{\eg}{\emph{e.g.}\xspace}

\newcommand{\secref}[1]{\mbox{Sec.~\ref{#1}}\xspace}

\newcommand{\figref}[1]{\mbox{Fig.~\ref{#1}}}

\newcommand{\tabref}[1]{\mbox{Table~\ref{#1}}}
\newcommand{\appref}[1]{\mbox{Appendix~\ref{#1}}}
\newcommand{\ignore}[1]{}
\usepackage{tikz}
\usepackage{pifont}
\usepackage{multirow}

\newcounter{note}[section]

\usepackage{amsmath}
\usepackage[mathscr]{eucal}
\usepackage{amsfonts}
\usepackage{authblk}

\newcommand{\AS}{{$\mathbb{A}_S$}\xspace} 
\newcommand{\AD}{{$\mathbb{A}_D$}\xspace} 
\newcommand{\BS}{{$\mathbb{B}_S$}\xspace} 
\newcommand{\BD}{{$\mathbb{B}_D$}\xspace} 
\newcommand{\CS}{{$\mathbb{C}_S$}\xspace} 
\newcommand{\CD}{{$\mathbb{C}_D$}\xspace} 
\newcommand{\TS}{{$\mathbb{T}_S$}\xspace} 
\newcommand{\TD}{{$\mathbb{T}_D$}\xspace} 
\newcommand{\BOS}{\textit{Server}\xspace} 

\begin{document}
\title{SoK: Security of Cross-chain Bridges: Characteristics, Attack Surfaces, Defenses, and Open Problems}
\title{SoK: Security of Cross-chain Bridges:  Attack Surfaces, Defenses, and Open Problems}

\author[1]{Mengya Zhang}
\author[2]{Xiaokuan Zhang}
\author[1]{Josh Barbee}
\author[3]{Yinqian Zhang}
\author[1]{Zhiqiang Lin}
\affil[1]{The Ohio State University}
\affil[2]{George Mason University}
\affil[3]{Southern University of Science and Technology}



    



\maketitle

\begin{abstract}
\noindent Cross-chain bridges are used to facilitate token and data exchanges across blockchains. Although bridges are becoming increasingly popular, they are still in their infancy and have been attacked multiple times recently, causing significant financial loss.  Although there are numerous reports online explaining each of the incidents on cross-chain bridges, they are scattered over the Internet, and there is no work that analyzes the security landscape of cross-chain bridges in a holistic manner. To fill the gap, in this paper, we performed a systematic study of cross-chain bridge security issues. First, we summarize the characteristics of existing cross-chain bridges, including their usages, verification mechanisms, communication models, and three categorizations. Based on these characteristics, we identify 12 potential attack vectors that attackers may exploit. Next, we introduce a taxonomy that categorizes cross-chain attacks in the past two years into 10 distinct types, and then provide explanations for each vulnerability type, accompanied by Solidity code examples. We also discuss existing and potential defenses, as well as open questions and future research directions on cross-chain bridges. We believe that this systematization can shed light on designing and implementing cross-chain bridges with higher security and, more importantly, facilitating future research on building a better cross-chain bridge ecosystem.
\looseness=-1
\end{abstract}

\section{Introduction}
\label{sec:intro}

Recent years have witnessed the rise in popularity of many widely used applications in blockchains, which allow users not only to simply transfer tokens, but also perform computations by taking advantages of the Turing-Complete smart contracts. 
In particular, smart contracts are self-executing programs that usually serve as the backends of Decentralized Applications (DApps) in blockchains (e.g., Ethereum~\cite{ethereum}).
DappRadar~\cite{dapp-radar} reports that, on average in 2022, all DApps combined had 2.37 million daily Unique Active Wallets (dUAW).
Despite the popularity of blockchain applications, they are restricted to a single blockchain domain;
blockchains cannot communicate or share data since they are isolated from each other. 
For example, \textit{ETH} is the native token of Ethereum and is limited to be used in Ethereum.
To use \textit{ETH} in other blockchains, users oftentimes need to sell the tokens to fiat currencies and trade them for tokens in the targeted blockchains. These steps are time-consuming, complex, and prone to risks.

To meet the increasing demand for conducting cross-chain transactions,
cross-chain bridges are created to facilitate communications between different blockchains.  
For example, AnySwap~\cite{anyswap} is a cross-chain bridging protocol that allows users to move assets 
between various blockchains such as Solana~\cite{solana} and Ethereum; 
ChainBridge~\cite{chainbridge} 
supports bridging between EVM and Substrate-based chains, where Substrate is a framework for creating custom blockchains~\cite{substrate}.
Currently, there are around 80 cross-chain bridges~\cite{dezentralizedfinance},
which support various user demands.

Due to their growing popularity and Total Value Locked (TVL) of cross-chain bridges,
bridges increasingly attract the attention of attackers.
For instance, the first reported attack on a cross-chain bridge was the 
ChainSwap hack~\cite{chainswap-attack} on Jul-10-2021, resulting in a loss of \$8 million. 
In the recent Nomad hack~\cite{nomad-attack} that occurred on Aug-02-2022,
the attacker stole around \$190 million worth of tokens. 
As reported by Chainalysis~\cite{chainalysis}, 
monetary losses caused by attacks on cross-chain bridges reached \$2 billion from January to July 2022, which represents 69\% of the total loss of all DeFi attacks in the same period. 

Despite the prosperity of cross-chain bridges and the severity of attacks, research on the security of cross-chain bridges is still in the early stage. 
One recent work~\cite{zhang2022xscope} focuses on detecting existing bridge attacks, while most existing works focus on measuring the interoperability between blockchains~\cite{buterin2016chain,belchior2021survey,wang2021sok}, summarizing current cross-chain protocols~\cite{zamyatin2021sok}, or proposing new protocols~\cite{shadab2020cross,wang2022efficient,liang2022privacy,su2022cross}.
There is only one SoK paper that analyzed eight attack surfaces from bridge components based on eight attacks~\cite{lee2022sok}.
However, they did not provide a holistic view of the security issues on cross-chain bridges,
such as a systematic characterization of cross-chain bridges, their vulnerabilities,
and all reported attacks. 
%
We provide a more comprehensive comparison with related work in~\secref{sec:related}.

\bheading{This paper.}
To clear the fog that surrounds the dark side of cross-chain bridges, this paper presents a systematic study on the security issues of cross-chain bridges. 
First, we present the characteristics of bridges by introducing four types of verification methods (\textit{external verification}, \textit{local verification}, \textit{optimistic verification}, and \textit{native verification}), two types of communication models (\textit{lock-and-mint;
burn-and-release} model and \textit{liquidity-pool-based} model), three categorization methodologies (\textit{trust mechanism}, \textit{layer connection}, and \textit{functionalities}), and usages (\textit{tokens trades}, \textit{governance}, \textit{lending and borrowing}, and \textit{staking}). 
We also summarize the top 30 cross-chain bridges according to their TVL on Aug-31-2023 from Chainspot~\cite{chainspot}. We find most bridges use the \textit{external verification} as their underlying protocol and deploy the \textit{lock-and-mint; burn-and-release} communication model.
Moreover, based on these characteristics, we examine the workflow of cross-chain bridges and identify 12 potential pitfalls that attackers might exploit. 

Next, we provide a taxonomy of existing bridge attacks by investigating all 31 bridge attacks from July 2021 to July 2023~\cite{certik,halborn} into 10 distinct types of vulnerabilities. 
The fundamental vulnerabilities that led to the attacks are categorized into four categories: \textit{permission issue (PI)}, \textit{logic issue (LI)}, \textit{event issue (EI)}, and \textit{front-end issues (FI)}. For each category of attacks, we give concrete examples in Solidity code.
Furthermore, we summarize existing defenses to these attacks and give recommendations to bridge developers for both on-chain backends (\ie, smart contracts) and off-chain servers\footnote{Cross-chain bridge server facilitates communication and coordination between different blockchains, ensuring that the assets are transferred securely and accurately.} to help them build more secure services. 
Finally, we discuss open problems in the cross-chain bridge ecosystem, and propose potential directions for future research. 


\bheading{Contributions.} Our contributions are:
\begin{packeditemize}
\item We give an overview of characteristics of cross-chain bridges, including verification mechanisms, communication models, and classifications based on different properties, including trust mechanism, layer connection, and functionalities (\secref{sec:pre});
\item We identify 12 potential attack surfaces based on the characteristics of cross-chain bridges (\secref{sec:vuls});
\item We provide a taxonomy on all the reported cross-chain bridge attacks (\secref{sec:taxonomy});
\item We summarize the existing defenses to the reported bridge attacks and provide recommendations to bridge developers (\secref{sec:defense}); 
\item We discuss open questions and future works on cross-chain bridge research (\secref{sec:question}).
\end{packeditemize}

\bheading{Scope.} The main focus of this study is to 1) summarize the characteristics of existing cross-chain bridges, 2) identify potential attack surfaces, 3) analyze all existing cross-chain bridge attacks happened in the wild, and 4) provide lessons learned from these incidents. 
While we discuss defense mechanisms in the paper, a systematic study on how to build mitigations or detection tools is not the goal of this paper.
A comprehensive study of cross-chain protocols (\eg, \cite{zamyatin2021sok,shadab2020cross,wang2022efficient,su2022cross,liang2022privacy}) is also out of the scope.
\section{Background}
\bheading{Blockchain} is a decentralized ledger built on top of a Peer-to-peer (P2P) network~\cite{p2p}
 consisting of distributed peers.
A blockchain can be considered as a continuously growing list of blocks.
Each block contains transactions that either transfer tokens in Bitcoin-like blockchains or perform more complex operations based on the computation logic in Ethereum-like blockchains. 

\bheading{Smart contracts} were first invented in Ethereum. They are programs that are stored and automatically executed according to the agreements between different parties~\cite{sc}, such as users and developers.
Once smart contracts are deployed to blockchains, they become immutable\footnote{One exception is the upgradable smart contract~\cite{upgrade-contract}, which can be updated.}.
Thus,  bugs in smart contracts are hard to fix due to their immutability.
While smart contracts can suffer from attacks, they have been the foundations of the Decentralized Finance ecosystem, and have been adopted to many widely used blockchains, 
such as BNB~\cite{bnb} and Fantom~\cite{fantom}. 

\bheading{Transactions} are data structures recording cryptocurrency information in the blockchains, which are either message calls to the smart contracts or simple token transfers to the blockchain users~\cite{tx}. 
Once transactions are mined to blockchains, the changes to the blockchain will be permanent and cannot be reverted.
Moreover, a transaction normally needs gas fee to pay for its computation cost. For example, Ethereum needs users to pay the transaction fee using the native token \textit{ETH}~\cite{fee}. In particular, the fee consists of gas used, \ie, the total units of the cost and gas price, \ie, the value users would like to pay for every unit. Furthermore, the higher the gas price is, the quicker a transaction usually is mined since miners are motivated to maximize their profits.

\bheading{Tokens} are digital assets defined in the applications that reside in the blockchains.
Except for the native tokens of blockchains (\eg, \textit{ETH} on Ethereum), tokens have their own  addresses. 
Moreover, every user can create his/her own token by implementing the smart contract standards required by blockchains, such as ERC20~\cite{erc20} and ERC721~\cite{erc721} in Ethereum.
Tokens can be mainly classified into three categories:
\begin{packeditemize}
    \item \textbf{Fungible tokens} are not unique, divisible, and interchangeable tokens that represent some values
    (similar to fiat currency such as USD). ERC20 tokens are one popular type of fungible tokens.
    \item \textbf{Non-fungible tokens} (NFTs) are non-divisible assets that are unique to each other. ERC721 tokens are one type of NFTs.
    \item \textbf{Wrapped tokens} represent cryptocurrencies pegged to the value of the related original tokens~\cite{wtoken}, including both fungible and non-fungible tokens. For example, \textit{WETH} is the wrapped version of \textit{ETH}. 
\end{packeditemize}

\bheading{DApps} are decentralized applications relying on blockchains~\cite{dapp}. Compared with traditional applications (e.g., websites or mobile apps), there is no difference in the frontends, but the backends of DApps are usually composed of multiple smart contracts in blockchains. In particular, cross-chain bridges are an increasingly popular type of DApps. 

\bheading{Liquidity pools} are the places that lock the tokens in smart contracts~\cite{lp}. 
Liquidity pools are designed to encourage users to provide liquidity by depositing tokens to the pools. Such users are called liquidity providers. Moreover, liquidity pools will distribute rewards to liquidity providers in terms of crypto tokens, based on the liquidity they have supplied. 
While liquidity pools facilitate token exchanges, liquidity providers are exposed to impermanent losses caused by the token price decrease, compared to the original price.

\section{Related Work}
\label{sec:related}
In this section, we discuss related works that study bridge attacks.
We also discuss related works on blockchain interoperability, cross-chain protocols, atomic swaps, and side chains.
Different from all the existing works, our work provides a systematization of attacks on cross-chain bridges, and further discusses open problems and future directions.

\begin{table}[]
\footnotesize
\resizebox{1.0\linewidth}{!}{
\setlength{\tabcolsep}{3pt}
\begin{tabular}{lllllllll}
\toprule
Paper & Research & \#Attack & \#Real-world  & \#Vul  & Code & Open \\
& Focus & Surface & Attack & Type & Example? & Problems? \\
\midrule
 Lee \etal~\cite{lee2022sok} & Bridge Component  & 8 & 8 & 8 & No & No\\
 {\bf This Paper} & Communication Model & 11 & 31 & 10 & Yes & Yes\\
 \bottomrule
\end{tabular}}
\caption{Comparison between our paper and Lee et al.~\cite{lee2022sok}}
\label{tab:prior}
\vspace{-13pt}
\end{table}

\bheading{Bridge attack SoK.}
As shown in \tabref{tab:prior}, Lee \etal~\cite{lee2022sok} conducted an analysis of bridge attacks from four components, including custodians, debt issuers, communicators, and token interfaces. They only summarized eight  possible attack surfaces based on eight real-world examples and discussed potential mitigation strategies. 
In contrast, our research takes a different approach. We systematically proposed 12 potential attack surfaces, drawing from two widely used bridge communication models. Our work offers a comprehensive taxonomy that encompasses reported 31 real-world cross-chain bridge attacks (from July 2021 to July 2023), which are classified into 10 vulnerability types. For each vulnerability, we provide source code examples, an explanation of related real-world attacks, and potential solutions.
We also discuss open problems.

\bheading{Bridge attack detection.} 
Zhang \etal~\cite{zhang2022xscope} introduce a tool designed to uncover security breaches, identify bridge attacks originating from transactions, and assess its effectiveness across four cross-chain bridges. Notably, the tool identifies vulnerabilities in cross-chain bridges, encompassing issues such as unrestricted deposit emissions, inconsistent event parsing, and unauthorized unlocking—elements integral to our condensed attack surface analysis. Our work primarily centers on categorizing real-world attacks and their associated attack surfaces, as opposed to presenting a methodology for their detection. Detection mechanisms are considered as potential avenues for future research in our work.

\bheading{Blockchain interoperability.}
The capability of different blockchains to communicate with each other is known as blockchain interoperability. According to \textit{Buterin} ~\cite{buterin2016chain}, there are three primary types of blockchain interoperability, which are centralized or multi-signature notary schemes, side chains or relays, and hash-locking. \textit{Belchior} \etal~\cite{belchior2021survey} conduct a literature review on blockchain interoperability and classify the studies into three categories: public connectors, blockchain of blockchains, and hybrid connectors. \textit{Wang}~\cite{wang2021sok} performs a systematic study on blockchain interoperability and classifies it into three types (chain-based, bridge-based, and DApp-based) based on current research. These works focus on measuring the current state of blockchain interoperability, while our work explores the applications of blockchain interoperability on cross-chain bridges. 

\bheading{Cross-chain protocols.}
Cross-chain protocols enable data sharing among multiple blockchains, similar to blockchain interoperability. 
\textit{Zamyatin et al.}~\cite{zamyatin2021sok} measure current cross-chain communication protocols and classify them into two categories: exchange protocols for token exchange and migration protocols for asset transfer between blockchains. 
\textit{Shadab et al.}~\cite{shadab2020cross} propose a uniform 3-phase protocol for general cross-chain transactions that complies with uniformity requirements and adds a third end-to-end constraint. 
\textit{Wang et al.}~\cite{wang2022efficient} design and implement a new cross-chain model based on cross-chain scheduling, while \textit{Su et al.}~\cite{su2022cross} propose an asynchronous cross-chain model that ensures cross-chain atomicity based on transaction dependence. 
\textit{Liang et al.}~\cite{liang2022privacy} propose a cross-chain privacy protection scheme to address identity privacy leakage.
These works measure or propose cross-chain protocols, while our work summarizes the cross-chain protocols used in current cross-chain bridges, such as CCIP from Chainlink.

\bheading{Atomic swaps.} 
Atomic swap can facilitate local verification, which is one type of four cross-chain bridge verification mechanisms (see~\secref{sec:pre:pro}). 
They do not require any intermediary party to enable the exchange. 
In 2018, \textit{Herlihy}~\cite{herlihy2018atomic} introduced the atomic swap protocol, which guarantees that the exchange is successfully completed or not executed at all. 
\textit{Thyagarajan et al.}~\cite{thyagarajan2022universal} proposes a universal protocol for swapping tokens across blockchains, utilizing adaptor signatures and time-lock puzzles. 
\textit{Gugger}~\cite{gugger2020bitcoin} describes a protocol for atomic swaps between Monero and Bitcoin, which can be generalized to similar blockchains.
\textit{Miraz et al.}~\cite{miraz2019atomic} discusses atomic swaps, including their workflow, current state, future trends, and possible applications. 
\textit{Borkowski et al.}~\cite{borkowski2018towards} presents an overview of the current state-of-the-art in cross-chain atomic swaps, as well as prominent blockchains and relevant ongoing and operational projects. 
Our work briefly mentions atomic swaps as a potential cross-chain bridge underlying protocol and classify them to the local verification categorization.

\bheading{Side chains.} 
Side chains can increase the main network’s scalability by processing and batching large amounts of transactions before submission on the main blockchain.
Pegged side chains were first introduced by \textit{Back et al.}~\cite{back2014enabling} in 2014, allowing for the transfer of assets between different blockchains, such as Bitcoin.
\textit{Singh et al.}~\cite{singh2020sidechain} provide a comprehensive review of various state-of-the-art side chains and platforms, analyzing their impact and identifying their limitations.
\textit{Kiayias et al.}~\cite{kiayias2020proof} propose the side chain construction that facilitates communication between Proof-of-Work blockchains without trusted intermediaries, while \textit{Gaži et al.}~\cite{gavzi2019proof}  developed a side chain construction suitable for Proof-of-Stake systems. 
These works focus on side chains primarily for scalability concerns, while our work concentrates on cross-chain bridges that enhance blockchain interoperability.

\section{Characterizing Cross-chain Bridges}
\label{sec:pre}
In this section, we first explain the common usages of bridges.
Second, we illustrate the four types of verification methods, along with some representative protocols. Third, we introduce the two prominent communication models in cross-chain bridges: \textit{lock-and-mint; burn-and-release} model and \textit{liquidity-pool-based} model. 
Furthermore, we categorize the bridges based on their trust mechanism, layer connection, and functionalities. 
We also summarize the top 30 bridges with describing their characteristics in~\tabref{tbl:bridge}, according to the TVL on August 31, 2023 from Chainspot~\cite{chainspot}. 

\subsection{Bridge Usages}
\label{sec:pre:usage}

Bridges exchange data between multiple blockchains, including messages and tokens. They are mainly used for token trades, as follows. We also describe other potential usages, including governance, lending and borrowing, and staking.
\tabref{tbl:usage} presents these usages. 

\begin{table}[t]
    \centering
    \scriptsize
    \begin{tabular}{lll} \toprule
    Classification & Usages & Example \\ \midrule
    Token transfer & token transfer & Anyswap \\
    Token withdraw & native token withdraw & Anyswap \\ 
    Token swap & token exchanges & Anyswap \\
    Governance & extend governance & Aave  \\
    Lending and borrowing & borrow and deposit & Cross-chain Loans\\ 
    Staking & stake and claim rewards & RADAR\\ 
    \bottomrule
    \end{tabular}
    \caption{\small Summary of potential bridge usages.}
    \label{tbl:usage}
    \vspace{-10pt}
\end{table}

\bheading{Cross-chain token transfer.} As shown in~\figref{figs:workflow}, \AS can submit a transaction \textit{TxLock} in Ethereum to transfer 10.0 \textit{ETH} to the bridge account, and then the bridge submits the transaction \textit{TxMint} to transfer wrapped tokens (10.0~\textit{WETH}) to \AD in BNB. 

\bheading{Cross-chain token withdraw.} Continuing with the above example, \AD submits a transaction \textit{TxBurn} in BNB and burns the wrapped tokens (10.0 ~\textit{WETH}). Then the bridge submits the related transaction \textit{TxRelease} to release  10.0 \textit{ETH} back to the user.

\bheading{Cross-chain token swap.} Similar to the token transfer, cross-chain token swap is another major usage of bridges. For instance, a user can swap 10.0 \textit{USDC} from Ethereum to BNB for 10.0 \textit{USDT}.

\bheading{Governance} is a popular way to manage and implement changes. In some DApps, users who hold the native governance tokens are allowed to submit and vote for proposals, to participate in the future of DApps. For instance, Aave~\cite{aave} holders of \textit{AAVE} and/or \textit{stkAAVE} tokens are given the rights of proposal and vote. 
To realize the governance across chains, Aave proposes governance bridges to extend its governance on Ethereum to other blockchains, such as Polygon~\cite{aave-bridge}. 

\bheading{Lending and borrowing} have always been fundamental services provided by DeFi platforms. Cross-chain lending and borrowing can bring flexibility to DeFi users. For example, Crosschain Loans~\cite{crosschain-loans} supports the cross-chain lending between EVM compatible blockchains. Users can borrow Ethereum tokens by depositing assets on other blockchains as collateral. 

\bheading{Staking} locks up a certain amount of assets, serves as a proof of work, and earns passive incomes. Previously, staking is limited to the single native blockchain. With cross-chain staking, users can stake across different blockchains and claim staking rewards on any blockchain, regardless of which blockchain the staked assets are deposited. For example, RADAR Cross-Chain Token Staking~\cite{radar} allows users to stake and claim rewards on any blockchain.

\subsection{Verification Mechanisms}
\label{sec:pre:pro}
Every bridge has its own unique protocol that is responsible for monitoring blockchains, achieving consensus, and relaying messages as necessary. Cross-chain bridge verification mechanisms can be classified into four categories based on the method of verifying cross-chain transactions: external verification, optimistic verification, local verification, and native verification~\cite{verification}.
\tabref{tbl:veri} presents four types of bridge verification mechanisms.

\begin{table}[t]
    \centering
    \scriptsize
    \begin{tabular}{lll} \toprule
    Classification & Verification Mechanism & Protocol \\ \midrule
    External Verification & external trusted validators & CCIP, LayerZero \\
    Optimistic Verification & honest watchers & Optics \\ 
    Local Verification & atomic swaps or smart contracts & Connext’s NXTP\\
    Native verification & a light client & MAP Protocol \\
    \bottomrule
    \end{tabular}
    \caption{\small Four types of bridge verification mechanisms.}
    \label{tbl:veri}
    \vspace{-10pt}
\end{table}

\begin{table*}[t]
    \centering
    \resizebox{1\linewidth}{!}{
    \footnotesize
    \begin{tabular}{rrrrrrrr} 
    \toprule
      \multirow{2}{*}{Bridge} & \multirow{2}{*}{TVL} & \multirow{2}{*}{Verification} & \multirow{2}{*}{Comm. Model} & \multicolumn{3}{c}{Classifications} & \multirow{2}{*}{Blockchains}\\
      \cline{5-7} 
       & & & & Trust Mechanism & Layer & Functionalities & \\
      \midrule	
      Polygon Bridge & \$2,370,560,407 & External & Lock-and-Mint & Semi-Trustless & L1 - L2 & Chain-Specific & Ethereum, Polygon \\  
      Arbitrum Bridge& \$1,507,382,649 & External & Lock-and-Mint & Trusted & L1 - L2 & Chain-Specific & Ethereum, Arbitrum \\
      Optimism Bridge& \$560,154,958 & External & Lock-and-Mint & Trusted & L1 - L2 & Chain-Specific & Ethereum, Optimism \\
      Portal Token Bridge& \$173,120,945 & External & Lock-and-Mint & Trusted & L1 - L2 & Chain-Specific & 30 blockchains~\cite{wormhole-chain} \\ 
      Orbit Bridge& \$161,048,342 & External & Lock-and-Mint & Trusted & L1 - L2 & Generalized & 21 blockchains~\cite{orbit-chain}\\ 
      Rainbow Bridge& \$89,017,728 & Native & Lock-and-Mint & Trustless & L1 - L1 & Chain-Specific & Ethereum, Near \\
      xDAI Bridge& \$73,462,737 & External & Lock-and-Mint & Trusted & L1 - L2 & Asset-Specific & Ethereum, Gnosis \\
      Satellite by Axelar & \$66,646,277 & Native & Lock-and-Mint & Trustless & L1 - L2 & Generalized & 14 blockchains~\cite{satellite-chain} \\ 
      Synapse Protocol & \$32,255,119 & External & Lock-and-Mint & Semi-Trustless & L1 - L2 & Application-Specific & 19 blockchains~\cite{synapse-19} \\
      Hop.Exchange & \$14,487,940 & Local & Lock-and-Mint & Trustless & L1 - L2 & Application-Specific & 7 blockchains~\cite{hop}\\
      Across Protocol & \$13,965,554 & Optimistic & Liquidity-Pool-Based & Trustless & L1 - L2 & Application-Specific & 5 blockchains~\cite{across-chain} \\
      Voltage Bridge& \$11,922,967 & Local & Liquidity-Pool-Based & Trustless & L1 - L2 & Chain-Specific & Ethereum, Fuse, Bnb\\
      Hot Cross Multi-Chain Bridge& \$7,098,673 & External & Lock-and-Mint & Trusted & L1 - L2 & Application-Specific & Ethereum, Avalanche, Bnb \\
      BoringDAO Bridge& \$6,900,409 & External & Lock-and-Mint & Trusted & L1 - L2 & Application-Specific & 14 blockhains~\cite{boringdao-chain}\\
      Multichain & \$6,898,501 & External & Lock-and-Mint & Trusted & L1 - L2 & Application-Specific & 26 blockhains~\cite{multichain-chain}\\
      ioTube Bridge& \$6,500,675 & External & Lock-and-Mint & Trusted & L1 - L2 & Chain-Specific & Ethereum, IoTeX, Bnb, Polygon\\
      ChainPort Bridge& \$6,166,914 & External & Lock-and-Mint & Trusted & L1 - L2 & Application-Specific & 17 blockchains~\cite{chainport-chain}\\
      Wrap Protocol & \$4,250,410 & External & Lock-and-Mint & Trusted & L1 - L1 & Chain-Specific & Ethereum, Tezzo \\ 
      Avalanche Bridge & \$4,126,385 & External & Lock-and-Mint & Trusted & L1 - L1 & Chain-Specific & Ethereum, Avalanche \\ 
      ThunderCore Bridge& \$4,025,853 & External & Lock-and-Mint & Trusted & L1 - L2 & Chain-Specific & 4 blockchains~\cite{thundercore-chain} \\ 
      RenBridge & \$3,386,159 & External & Lock-and-Mint & Trusted & L1 - L1 & Application-Specific & Ethereum, Bitcoin \\
      Cross-Chain Bridge & \$2,901,984 & External & Liquidity-Pool-Based & Trusted & L1 - L2 & Application-Specific & 5 blockchains~\cite{cross-chain} \\
      Sovryn Bridge & \$2,753,302 & External & Liquidity-Pool-Based & Trusted & L1 - L2 & Chain-Specific & Ethereum, RSK, Bnb \\
      Hyphen Bridge& \$2,753,037 & External & Liquidity-Pool-Based & Trusted & L1 - L2 & Application-Specific & 7 blockchains~\cite{hyphen-chain}\\
      Celer cBridge & \$2,735,285 & Native & Lock-and-Mint & Trustless & L1 - L2 & Application-Specific & 34 blockchains~\cite{cbridge-chain}\\
      RSK Token Bridge & \$2,965,215 & External & Lock-and-Mint & Trusted & L1 - L2 & Chain-Specific & Ethereum, RSK \\
      Celo Optics Bridge & \$1,522,932 & Optimistic & Lock-and-Mint & Trustless & L1 - L2 & Generalized & Ethereum, Polygon, Celo \\
      Nomad & \$1,244,025 & Optimistic & Lock-and-Mint & Trustless & L1 - L2 & Application-Specific & 4 blockchains~\cite{nomad-chain} \\
      Allbridge & \$836,557 & External & Lock-and-Mint & Trusted & L1 - L2 & Application-Specific & 21 blockchains~\cite{allbridge-chain} \\
      SOY Bridge & \$740,987 & External & Lock-and-Mint & Trusted & L1 - L2 & Application-Specific & 5 blockchains~\cite{soy-chain}\\ 
     \bottomrule
    \end{tabular}
    }
    \caption{\small Summary of 30 popular cross-chain bridges, ranked by the Total Locked Value (TVL) on August 31, 2023 from Chainspot~\cite{chainspot}. 
    \textit{Lock-and-Mint} represents the \textit{Lock-and-Mint; Burn-and-Release} communication model. 
    }
    \label{tbl:bridge}
    \vspace{-10pt}
\end{table*}

\subsubsection{External Verification}
External Verification (EV) relies on a single or set of external validators, who are responsible for verifying the validity of cross-chain transactions. Bridges operating under this categorization assume that \textit{most validators are honest}, and typically adopt multi-signature or multi-party computation (MPC). With the multi-signature mechanism, a threshold is set for the number of validators required to sign the transaction, and each validator has a complete private key. In contrast, MPC requires the validators to jointly generate a private key. Despite being simple to implement and widely used, this mechanism is exposed to considerable risks if the external validators are not trusted or are hacked. For instance, if the private keys of these validators are compromised, the bridges may be taken over by attackers. We illustrate two protocols, including Chainlink's cross-chain interoperability protocol (CCIP) and an omnichain interoperability protocol (LayerZero).

\bheading{Cross-Chain Interoperability Protocol} (CCIP) facilitates standardized and secure communication and information exchange among various blockchain networks. To develop their CCIP, Chainlink utilizes Decentralized Oracle Networks (DONs) and the Anti-Fraud Network~\cite{chainlink-ccip}. DONs provide off-chain services and data, such as random number generation, and enable secure validation of cross-chain transactions through the consensus of multiple oracle nodes. The Anti-Fraud Network, on the other hand, detects and halts malicious cross-chain activity to safeguard users. Chainlink's CCIP supports various cross-chain services, such as programmable token bridges, cross-chain bridges, and cross-chain DApps. Specifically, Chainlink's CCIP enables a smart contract on the source blockchain to use their messaging router to send a secure message to a destination chain's messaging router. The destination messaging router then validates and forwards the message to the destination smart contract.

\bheading{Omnichain interoperability protocol} (LayerZero) facilitates seamless communication between multiple blockchains. Unlike other protocols, Omnichain has no wrapping or intermediary layers. 
LayerZero is one of such protocols~\cite{zarick2021layerzero}, which offers trustless messaging and can operate on any blockchain network, such as Ethereum and Bitcoin. 
LayerZero leverages an Oracle and a Relayer to securely pass messages between different domains off-chain. The Oracle sends a block header from the source blockchain to the destination blockchain, while the Relayer passes a transaction proof. If the block header and proof match and are validated, the cross-chain message is sent to the destination blockchain. In practice, LayerZero leverages the Chainlink's DONs to ensure trustless delivery of messages between disparate chains, as an Oracle. For example, Stargate~\cite{stargate} is a fully composable cross-chain bridge protocol built on LayerZero that enables native asset transfers between different blockchain networks.

\subsubsection{Optimistic Verification}
Optimistic Verification (OV) operates on the assumption that a transaction is valid and initiates a challenge period to verify its accuracy. If the transaction is found to be invalid during the challenge period, the transaction will be reversed, and malicious actors will be penalized. However, bridges under this categorization often have a longer latency due to the challenge period. Moreover, such bridges \textit{require at least one honest watcher} to verify updates and detect fraudulent activity. We illustrate an optimistic inter-chain communication protocol.

\bheading{Optimistic Interchain Communication Protocol} (Optics) is a protocol that enables cost-efficient message transfer between consensus systems~\cite{optics}. It relies on channels to send messages as raw bytes from one chain to another. Once a channel is established, any application on a chain can use it to send messages to any other chain, without requiring a light client or extra gas for verifying remote chain block headers. Optics contains initial implementations of on-chain contracts in Solidity code and off-chain system agents in Rust code.
Applications that use Optics need to implement their messaging protocol and are facilitated by Router contracts. For instance, Nomad~\cite{nomad} extends the Optics protocol. During a 30-minute window, challengers can question the validity of transactions.

\subsubsection{Local Verification}
Local Verification (LV) requires only the parties involved in the transaction to verify it, making it a two-party verification mechanism that maintains the security of the underlying blockchain. This mechanism \textit{inherits the security of the underlying blockchain} and is popularly used for atomic swaps, which ensure transaction atomicity using Hash Time Lock Contracts (HTLC). HTLC uses hashlock and timelock to lock assets within a time window and requires consensus from both parties to complete the payment. The receiver of the payment can either accept it before it expires or forfeit it. However, LV has limitations and cannot be easily extended to other applications in addition to token transfers. We illustrate Connext's NXTP.

\bheading{Connext's NXTP} provides a straightforward method for cross-chain transfers and contract calls that is fully noncustodial~\cite{nxtp}. The protocol involves three phases: route selection, preparation, and fulfillment. During route selection, routers bid to complete the transfer. In the preparation phase, users initiate a transaction on the source blockchain through the NXTP contracts. In the fulfillment phase, the message is signed to complete the transaction on the destination blockchain. At any point, either party can cancel the transaction if necessary. Once the transaction has expired, both parties can cancel the transaction.

\begin{figure*}[h]
\centering
\includegraphics[width=1\linewidth]{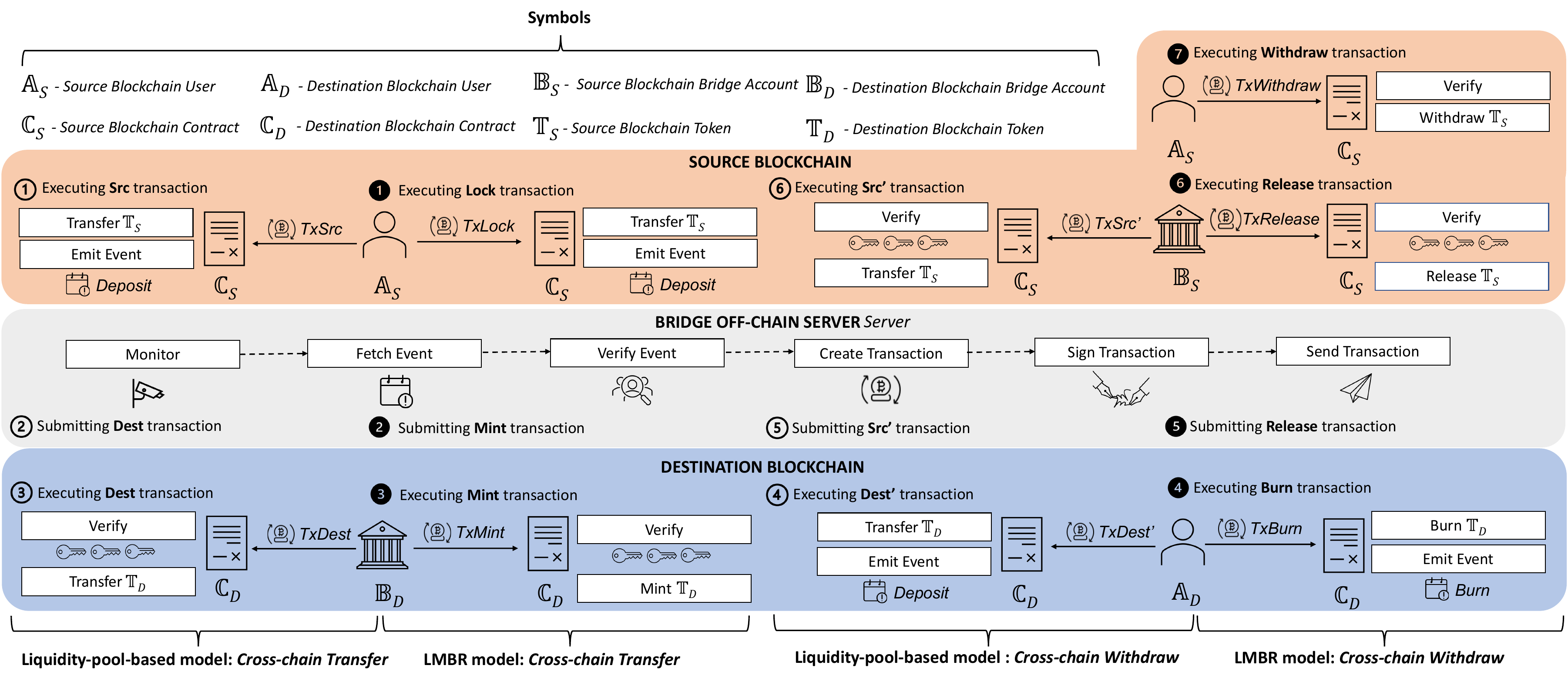}
\caption{\small The workflows of two cross-chain bridge communication models.
\cc{Lock-and-Mint; Burn-and-Release (LMBR)} model consists of \ding{182} - \ding{188}. 
 \textit{Cross-chain transfer}: \ding{182} User \AS submits a transaction \textit{TxLock} to \CS that transfers tokens \TS and emits \textit{Deposit} events; \ding{183} \BOS fetches and verifies the event for submitting \textit{TxMint} to \CD; \ding{184} \textit{TxMint} sent from \BD transfers the minted tokens \TD. 
 \textit{Cross-chain withdraw}: \ding{185} \AD submits the transaction \textit{TxBurn} to burn \TD; \ding{186} Similarly, \BOS fetches and verifies the \textit{Burn} event for submitting \textit{TxRelease} sent from \BS to release the locked tokens \TS in step \ding{187}; \ding{188} \AS finally submits \textit{TxWithdraw} to take the tokens back. 
\cc{Liquidity-pool-based} model consists of \ding{172} - \ding{177}.
 \textit{Cross-chain transfer}: \ding{172} Similarly, User \AS submits the transaction \textit{TxSrc} to \CS to deposit tokens \TS; \ding{173} \BOS submits the transaction \textit{TxDest} to \CD; \ding{174} \textit{TxDest} sent from \BD transfers the tokens \TD (token of the destination blockchain) from the liquidity pool to \AD. 
 \textit{Cross-chain withdraw}: \ding{175} User \AD sends tokens \TD back to \CD; \ding{176} \BOS submits the related transaction \textit{TxSrc'}; \ding{177} \BS transfers tokens \TS to user \AS. 
 }
\label{figs:workflow}
\vspace{-6pt}
\end{figure*}

\subsubsection{Native verification} 
Native verification (NV) requires the destination blockchain to run a light client of the source blockchain to verify the related events or states. 
Light clients store the minimum information required to verify the status change of blockchains in a self-verifiable manner. They keep block headers to confirm cross-chain messages and stay updated with any changes in blockchains to verify the relayed cross-chain transactions based on stored state information. Although this mechanism is the most secure, as it \textit{has the same security assumptions as the public chain}, it is expensive to maintain the light clients for all supported blockchains for bridges. Moreover, we illustrate one MAP Protocol. 

\bheading{MAP Protocol} is an interoperability protocol that facilitates cross-chain transactions, supported by the dedicated MAP blockchain~\cite{map}. The protocol consists of three layers: the MAP Protocol Layer, the MAPO Services Layer, and the MAPO Application Layer. The MAP Protocol Layer is responsible for verifying cross-chain assets and data through independent, self-verifying light clients deployed on each blockchain. The MAPO Services Layer serves as the execution layer for assets and data and supports developers to build applications for the MAPO Application Layer. Overall, MAP Protocol is designed to enable DeFi and other applications to reach their full potential while ensuring the safety of cross-chain assets.

Note in~\tabref{tbl:bridge}, the majority of bridges employ external verification (22), while a small number opt for optimistic verification (3), native verification (3), or local verification (2).

\subsection{Communication Models}
\label{sec:pri:mec}
There are two models commonly used by the bridges for providing cross-chain services, which are (1) {\bf lock-and-mint; burn-and-release} model and (2) {\bf liquidity-pool-based} model, respectively. 
Specifically, model (1) will mint new wrapped tokens (e.g., \textit{wBTC}) in the destination chain, which will be burnt to release tokens locked in the source blockchain, while model (2) only includes native assets (e.g., \textit{BTC}). 
To make it clear, we illustrate the workflow of these two models with common usages: cross-chain transfer and withdraw in~\figref{figs:workflow}. 

\subsubsection{Lock-and-Mint; Burn-and-Release Model}
In~\figref{figs:workflow}, Step \ding{182} to \ding{184} are the \textit{cross-chain transfer} and Step \ding{185} to \ding{188} are the \textit{cross-chain withdraw}, which is the reverse operation to the \textit{cross-chain transfer}.

\iheading{Cross-chain transfer.}
Assume user \AS wants to send money from one chain to his/her own account \AD in another chain, which is usually implemented via the \textit{cross-chain transfer}:

\iheading{\ding{182} Executing \textbf{Lock} transaction.} The first step is to submit a transaction \textit{TxLock} from \AS to \CS, performing a token transfer from \AS to the bridge account. Meanwhile, the \textit{Deposit} event is emitted for \BOS to listen and record. 

\iheading{\ding{183} Submitting \textbf{Mint} transaction.} In step two, the \textit{Deposit} event containing detailed information (e.g., token amount) is verified by \BOS. If the event is parsed and verified to be correct (e.g., emitted from the \CS), the related transaction \textit{TxMint} will be created, signed by validators, and then sent to call \CD.

\iheading{\ding{184} Executing \textbf{Mint} transaction.} In step three, \textit{TxMint} is sent from \BD. During the execution, the signatures from validators are verified. After the verification, related tokens \TD (wrapped tokens of \TS) are minted. In particular, the tokens are transferred from the bridge account or the \textit{zero} address~\cite{0x00000000} to \AD, which is appointed by \AS in \textit{TxLock}.

\iheading{Cross-chain withdraw.}
When user \AS needs to take his/her tokens that are locked in the source 
blockchain back, 
\AS conducts a \textit{cross-chain withdraw}.
The cross-chain withdraw can only be used after the cross-chain transfer.

\iheading{\ding{185} Executing \textbf{Burn} transaction.} In step four, \AD should submit a transaction \textit{TxBurn} calling \CD, to burn tokens \TD, by sending the tokens to the bridge account or the \textit{zero} address. Meanwhile, the \textit{Burn} event is emitted from \CD for \BOS to monitor. 

\iheading{\ding{186} Submitting \textbf{Release} transaction.} This step is similar to step \ding{183}. 
If the \textit{Burn} event is verified, the transaction \textit{TxRelease} will be created and signed by enough validators.

\iheading{\ding{187} Executing \textbf{Release} transaction.} In step six, \textbf{Release} is sent from \BS, the signatures from bridge validators are verified in \CS, and the tokens locked before will be released, after the verification. However, the tokens are not sent directly back to \AS yet. Instead, a proof showing that the tokens are available to withdraw is sent to \AS.

\iheading{\ding{188} Executing \textbf{Withdraw} transaction.} In step seven, \AS submits the transaction \textit{TxWithdraw} to \CS along with the proof, to withdraw the tokens. If the proof is valid, the bridge will transfer tokens back to \AS.

\subsubsection{Liquidity-Pool-Based Model}
For the liquidity-pool-based model in~\figref{figs:workflow}, Step \ding{172} to \ding{174} are the \textit{cross-chain transfer} and Step \ding{175} to \ding{177} are the \textit{cross-chain withdraw}.

\iheading{Cross-chain transfer.}
Assume user \AS wants to send money from one chain to his/her own account \AD in another chain, which is usually implemented via the \textit{cross-chain transfer}:

\iheading{\ding{172} Executing \textbf{Src} transaction.} 
To begin with, the bridge creates the liquidity pool that only contains token \TS. Different from normal liquidity pools, cross-chain bridge liquidity pools only have single type of tokens, to prevent impermanent losses for liquidity providers. 
Then, the liquidity providers add liquidity to the pool by depositing tokens \TS.
After preparations of liquidity pool, the first step is to execute a transaction \textit{TxSrc} from \AS to \CS, performing a token transfer and emitting the \textit{Deposit} event for \BOS to listen and record.

\iheading{\ding{173} Submitting \textbf{Dest} transaction.} This step is similar to the step \ding{183} in Lock-and-Mint; Burn-and-Release model. 

\iheading{\ding{174} Executing \textbf{Dest} transaction.} Before step three, the liquidity pools in the destination blockchain should be prepared as well. Similarly, the pool only containing tokens \TD is created by the bridge and the liquidity providers add liquidity to the pool. In particular, tokens \TS and \TD represent the same tokens in different blockchains. Meanwhile, the transaction \textit{Dest} sent from \BD to \CD will perform the verification and then transfer the tokens \TD from the liquidity pool to \AD.

\iheading{Cross-chain withdraw.}
\textit{Cross-chain withdraw} is similar to \textit{Cross-chain transfer}, except that the direction (from destination to source) is opposite from the direction of \textit{Cross-chain transfer} (from source to destination).

\iheading{\ding{175} Executing \textbf{Dest'} transaction.} To withdraw the tokens deposited in the source blockchain, user \AD initializes a transaction \textit{TxDest'} that sends the tokens \TD back to the bridge \CD. 

\iheading{\ding{176} Submitting \textbf{Src'} transaction.} This step is also similar to the step \ding{183} in Lock-and-Mint; Burn-and-Release model.

\iheading{\ding{177} Executing \textbf{Src'} transaction.} Transaction \textit{TxSrc'} is sent to \CS, verifying the data (e.g., proof) and transferring tokens \TS back to \AS then.

Note in~\tabref{tbl:bridge}, 25 bridges use \textit{lock-and-mint; burn-and-release} communication model and 5 bridges use \textit{Liquidity-Pool-Based} communication model. 

\subsection{Categorization}
\label{sec:pre:taxonomy}
We classify the cross-chain bridges based on different standards, including 1) trust mechanisms, 2) layer connections, and 3) functionalities.

\subsubsection{Trust Mechanism Classification}
Considering the trust mechanisms, we classify the bridges into three categories: trusted, semi-trustless, and trustless.


\iheading{Trusted bridges} typically have to trust a single or several validators, to perform related operations (e.g., signing and submitting transactions). 
The security relies on the reputations of these validators since they do not deposit collateral as penalty. For instance, the bridge Muiltichain~\cite{anyswap} is powered by Fusion’s DCRM technology~\cite{fusion} to ensure the security. 

\iheading{Semi-trustless bridges} require the validators to deposit assets for punishment if anything goes wrong. Such bridges may suffer hacks from the inappropriate operations from validators. For instance, if the private keys of enough validators are compromised, the bridge will be under attack. Moreover, the deposited assets from validators are required. 
Taking Synapse~\cite{synapse} as an example, it is
secured by cross-chain multi-party computation (MPC). In particular, if two-thirds of all validators sign the transactions, the consensus will be achieved and the further related transactions will be submitted to blockchains. 
\looseness=-1

\iheading{Trustless bridges} have the highest security, equivalent to the underlying blockchains, in which anyone can become a validator. 
These bridges usually perform operations using smart contracts or algorithms. 
For example, Celer cBridge~\cite{cbridge} depends on the decentralized State Guardian Network (SGN)~\cite{cbridge-trust} to monitor transactions and faithfully perform operations, such as passing the information and storing the locked tokens. In partciular, SGN is based on the Proof-of-Stake (PoS) consensus mechanism, which requires validators to stake some tokens to participate in the network.

According to~\tabref{tbl:bridge}, among the top 30 bridges, 20 are trusted, 2 are semi-trustless, and 8 are trustless.

\subsubsection{Layer Connection Classification.}
Blockchains can be divided into layer 1 (L1) and layer 2 (L2). L1 is the base blockchain protocol, such as Ethereum. L2 represents a third-party solution integrated with L1, allowing for better scalability, such as Polygon~\cite{polygon}. Considering the connections of different blockchains, bridges can also be classified into two categories: L1 - L1 and L1 - L2.


\iheading{L1 - L1 bridges} connect different L1 blockchains. For example, Near Rainbow Bridge~\cite{near-bridge} connects Ethereum and Near. 

\iheading{L1 - L2 bridges} establish a connection between L1 and various L2 blockchains, while L2 blockchains are also interconnected with each other. Anyswap~\cite{anyswap}, for instance, facilitates connectivity between 35 blockchains, including Ethereum, Polygon, and others.

Note in~\tabref{tbl:bridge}, among the top 30 bridges, only 4 bridges are L1 - L1 and 26 bridges are L1 - L2. 

\subsubsection{Bridge Functionality Classification.}
Based on functionality, bridges can be classified into asset-specific, chain-specific, application-specific, and generalized bridges.


\iheading{Asset-specific bridges} only allow a specific asset transfer. For example, xDAI bridge~\cite{xdai} supports \textit{DAI} and \textit{xDAI} transfer between Ethereum and Gnosis.

\iheading{Chain-specific bridges} are dedicated to certain blockchains. These bridges are usually developed by major blockchains, such as Polygon Bridge~\cite{polygon-bridge}, which allows users to transfer assets between Ethereum and Polygon.

\iheading{Application-specific bridges} have limited functionality and are designed to handle only specific applications, particularly asset exchanges. For example, AnySwap~\cite{anyswap} does not support transmitting state data or validation information.

\iheading{Generalized bridges} enable the transfer of various information types, such as assets, contract calls, proofs, and states. They are not restricted to particular blockchains or applications, and can instead facilitate communication between a diverse set of disconnected networks. 
For example, Chainlink~\cite{chainlink} supports cross-chain services, including the Programmable Token Bridge, other bridge implementations, and even cross-chain DApps.

Note in~\tabref{tbl:bridge}, among the top 30 bridges, one is asset-specific, 12 are chain-specific, 14 are application-specific, and three are generalized bridges. 

\section{Bridge Attack Surfaces}
\label{sec:vuls}

In this section, we illustrate 12 possible attack surfaces of cross-chain bridges, according to the two bridge models in \secref{sec:pri:mec}, as shown in~\tabref{tbl:surf}.
Initially, we outline the shared attack surfaces that exist in both communication models, followed by an introduction to the distinctive attack surfaces. 

\begin{table}[!htbp]
    \centering
    \footnotesize
    \resizebox{1.0\linewidth}{!}{
    \setlength{\tabcolsep}{3pt}
    \begin{tabular}{llll} 
      \toprule
      Models & Attack Surface & Step/Target & Example \\
      \midrule
      Both & A1 (Front-end phishing) & \BOS & deceiving users\\
           & A2 (Inaccurate deposit) & \ding{182}; \ding{172}, or \ding{175} & incorrect bridge account \\
           & A3 (Mishandling events) & \ding{183} or \ding{186}; \ding{173} or \ding{176} & incorrect Deposit event \\
           & A4 (Mismatched transactions) & \ding{183} or \ding{186}; \ding{173} or \ding{176} & \textit{TxMint} not sent \\
           & A5 (Single points of failures) & \BOS & private key leakage \\
           & A6 (Rugpull) & Bridge & bridge running away \\
           & A7 (Vulnerable Contracts) & \CS or \CD & reentrancy \\
      \cmidrule{1-4}
      M1 & A8 (Problematic mint) & \ding{184} & invalid mint permission \\ 
           & A9 (Fake burn) & \ding{185} & invalid burn approval \\
           & A10 (Incorrect release) & \ding{186} & invalid verification \\
           & A11 (Replayed withdraw) & \ding{187} & problematic proof \\
      \cmidrule{1-4}
      M2 & A12 (Inconsistent transfer) & \ding{174} or \ding{177} & incorrect transfer amount \\
      \bottomrule
    \end{tabular}}
    \caption{\small Summary of 12 possible attack surfaces, according to two bridge models.
        \textit{M1}: Lock-and-Mint; Burn-and-Release Model.
        \textit{M2}: Liquidity-Pool-Based Model.}
    \label{tbl:surf}
    \vspace{-10pt}
\end{table}

\subsection{Common Attack Surfaces} 
A1-A7 can be applied to both the lock-and-mint; burn-and-release model and liquidity-pool-based model. 

\bheading{A1: Front-end phishing.} Regardless of the communication model used, this type of attack is possible in \BOS. If an attacker gains access to the front-end of bridges, they can deceive users into transferring their funds, resulting in the theft of user funds.

\bheading{A2: Inaccurate deposit.} This attack can happen in step \ding{182}, \ding{172}, or \ding{175} in~\figref{figs:workflow}. Users need to deposit tokens to the bridge account. If the bridge account address is set incorrectly, the deposited tokens will be lost. Moreover, if the related events (e.g., \textit{Deposit}) are constructed and emitted incorrectly, users will not receive the tokens in the destination blockchain. 

\bheading{A3: Mishandling events.} This attack can occur in step \ding{183}, \ding{186}, \ding{173}, or \ding{176} in~\figref{figs:workflow}. \BOS will fetch and parse the events from the transactions. If these events are handled incorrectly, bridges might perform inappropriate actions, such as sending more money to users than desired. For example, if a \textit{Deposit} event recording a 10 \textit{ETH} transfer in the source blockchain is mistakenly treated as a 100 \textit{ETH} transfer, bridges will send 100 wrapped \textit{ETH} to the user in the destination blockchain. 

\bheading{A4: Mismatched transactions.} This attack can occur in step \ding{183}, \ding{186}, \ding{173}, or \ding{176} in~\figref{figs:workflow}. Cross-chain bridges are heavily relying on the off-chain bridge server (\BOS), for monitoring, parsing information, and submitting transactions. If the related transactions (e.g., \textit{TxMint} to mint wrapped tokens for users) are not sent at all, user will not receive their tokens. Moreover, if the transactions are constructed incorrectly (e.g., the receiver address is given with a wrong value), attacks can also happen. 

\bheading{A5: Single points of failures.} This attack can occur in \BOS in~\figref{figs:workflow}. Some bridges rely on external validators that can be compromised or engage in malicious behaviors, as follows. 
\begin{packeditemize}
    \item Compromised bridge validators. If the private keys of enough validators are compromised, attackers can take over the bridges to drain all the assets or mint tokens without limitations. 
    \item Malicious validators. Validators might be dishonest if the values in the bridges are far more than the values validators stake, as punishments for behaving maliciously.
\end{packeditemize}

\bheading{A6: Rugpull.} This attack can occur if the bridge itself is the attacker. Rugpulls are scams
in which the token founders suddenly vanish
and drain all the assets.
For example, users of Luna Yield, a farming project on the Solana blockchain, suffered from the rugpull attack on Aug-19-2021, in which the creators suddenly deleted the website and disappeared~\cite{rugpull-luna}. 
Additionally, the investors of a NFT game DApp, BlockVerse, withdrew all the money and disappeared on Jan-25-2022~\cite{rugpull-blockverse}.
Users should watch out for the possibility of rugpulls in cross-chain bridges, especially on newly created bridge projects. 

\bheading{A7: Vulnerable bridge smart contracts.} This attack can happen in \CS or \CD in~\figref{figs:workflow}. Common vulnerabilities in smart contracts (e.g., access control) are possible in bridge smart contracts. 
We illustrate some potential attacks, taking the liquidity pool smart contract code as an example (see ~\appref{app:surf}), including incorrect initialization, inappropriate function permission, unchecked balance, miscalculated token price, and inconsistent event. 
Existing vulnerabilities in DeFi can also exist in cross-chain bridges. Current cross-chain bridges only support simple financial services, which are mainly token transfers. However, we believe the attacking surface is much larger than the existing bridge attacks. 
\begin{packeditemize}
    \item Price manipulation attack. If the bridge relies on unstable price oracles that could be maliciously manipulated by attackers, the bridges will suffer from price manipulation attacks.
    \item Reentrancy attack. Reentrancy attack happens when the victim contract call a malicious contract who calls back to the original function in the victim contract, to steal money. In both functions \textit{addLiquidity} and \textit{removeLiquidity}, the reentrancy attacks occur. 
\end{packeditemize}

\subsection{Unique Attack Surfaces}
\subsubsection{Lock-and-Mint; Burn-and-Release Model}
A8-A11 are only applicable to the lock-and-mint; burn-and-release model. 

\bheading{A8: Problematic mint.} This attack can happen in step \ding{184} in~\figref{figs:workflow}. Once the tokens are locked in the source blockchain, the bridges will mint wrapped tokens in the destination blockchain. However, if the proof is not validated appropriately, attackers can mint more tokens than desired. 

\bheading{A9: Fake burn.} This attack can happen in step \ding{185} in~\figref{figs:workflow}. To get the tokens back in the source blockchain, users should burn the wrapped tokens in the destination blockchain. If there are some permission check issues that allow users to pass the burn, attacks can happen due to fake burns in which attackers burn nothing but get burn proofs. Moreover, if the \textit{Burn} event is not emitted correctly, there will be problems.

\bheading{A10: Incorrect release.} This attack can occur in step \ding{187} in~\figref{figs:workflow}. The related data need to be verified before creating the release proof. If the verification process is not handled correctly, the release proofs might be forged by attackers.

\bheading{A11: Replayed/Unlimited withdraw.} This attack can happen in step \ding{188} in~\figref{figs:workflow}. The release proof needs validation before allowing users to withdraw tokens. However, if the proof can be used multiple times or forged, users can steal tokens until the assets are drained. 

\subsubsection{Liquidity-Pool-Based Model}
{A12} is only applicable to the liquidity-pool-based model.  \looseness=-1

\bheading{A12: Inconsistent cross-chain transfer}. 
This attack can occur in step \ding{174} or \ding{177} in~\figref{figs:workflow}.
Assume that a user \AS transfers \textit{ETH} tokens from Ethereum (source blockchain) to BNB (destination blockchain). It is possible to suffer from the inconsistencies of cross-chain token transfer, in which the amount of transferred token in \textit{TxDest} or \textit{TxDest'} is incorrect (e.g., less than expected).

\section{A Taxonomy of Cross-chain Bridge Attacks and Vulnerabilities}
\label{sec:taxonomy}
In this section, we systematically categorize reported cross-chain bridge attacks and vulnerabilities into four categories, including Permission Issue (PI), Logic Issue (LI), Event Issue (EI), and Front-end Issue (FI), and summarize them in~\tabref{tbl:attack-class}. 
We also introduce the attacks caused by front-end phishing and replay attacks. 
In particular, we summarize the knowledge of cross-chain bridge attacks from reliable resources, including official bridge documentations (e.g., Wormhole~\cite{wormhole}), attack reports from blockchain companies (e.g., Certik~\cite{certik}), and blogs or posts about cross-chain bridges from technical organizations (e.g., HALBORN~\cite{halborn}). 
 \looseness=-1
 
\begin{table*}[t]
    \centering
    \resizebox{1.0\linewidth}{!}{
    \footnotesize
    \setlength{\tabcolsep}{2pt}
    \begin{tabular}{llllllll} 
      \toprule
      Type & Sub-type & Attack Surface & Victim Bridge & Attack Date & Loss (\$million) & Victim Blockchains & Defense \\
      \midrule
      PI & V1. Unchecked intermediary permission & A7($\mathbb{C}_S$ or $\mathbb{C}_D$) & Poly Network & Aug-10-2021  & 600 & Ethereum & Au;BB \\\cmidrule{2-8}
                & V2. Misused proof permission & A11($\mathbb{C}_S$) & Plasma & Oct-05-2021  & 850 & Polygon & BB \\
                        && A10($\mathbb{C}_S$ or $\mathbb{C}_D$) & Binance Bridge & Oct-06-2022  & 566 & Bnb & BB;HF \\\cmidrule{2-8}
                & V3. Problematic approval permission & A7($\mathbb{C}_S$ or $\mathbb{C}_D$) & Multichain & Jan-18-2022 & 3 & Ethereum & UVC;RIA \\
                                        &&& LI.FI & May-20-2022 & 0.6 & Ethereum & UVC \\
                                        &&& Rubic & Dec-25-2022 & 1.4 & Ethereum & UVC, Au \\\cmidrule{2-8}
                & V4. Invalid signature permission & A8($\mathbb{C}_D$)  & ChainSwap & Jul-10-2021 & 8 & Ethereum & Au;BB\\
                             &&& Wormhole & Feb-02-2022 & 320 & Solana, Ethereum & BB\\
                             &&& Multichain & Feb-15-2023 & 0.1 & Ethereum & UVC \\ \cmidrule{2-8}
                & V5. Leaked key permission & A5(\BOS) & WonderHero & Apr-07-2022 & 0.3 & Bnb & Au;UVC;BB\\ 
                           &&& Ronin Network & May-29-2022  & 625 & Ethereum & Au;IVD\\
                           &&& Horizon Bridge & Jun-24-2022  & 100 & Ethereum, Bnb & Au \\
                           &&& QAN Platform Bridge & Oct-11-2022  & 2 & Ethereum, Bnb & Au \\
                           &&& Rubic & Nov-02-2022  & 1.2 & Ethereum & Au \\
                           &&& pNetwork & Nov-04-2022  & 108 & Bnb & UVC \\
                           &&& Poly Network & Jul-01-2023 & 10.2 & Ethereum & UVC \\
                           &&& Multichain & Jul-06-2023 & 126 & Ethereum et al. & N/A \\
                           \cmidrule{1-8}
      
      LI & V6. Incorrect balance logic & A7($\mathbb{C}_S$ or $\mathbb{C}_D$)& Polkabridge  & Nov-22-2021 & 0.6 & Ethereum & UVC \\
                &&& Poolz Finance & Mar-15-2023 & 0.6 & Ethereum, Bnb, Polygon & Au \\
                & V7. Inaccurate initialization logic & A7($\mathbb{C}_S$ or $\mathbb{C}_D$) & Nomad & Aug-02-2022  & 190 & Ethereum, Moonbeam & BB\\
                &&& Allbridge & Apr-01-2023 & 0.57 & Bnb & UVC \\
      \cmidrule{1-8}
      EI & V8. Incorrect event emission & A2($\mathbb{C}_S$) & Qubit & Jan-27-2022 & 80 & Ethereum, Bnb & BB\\
                                    &&& Meter.io & Feb-05-2022 & 4.2 & Ethereum, Bnb, Moonriver & BB \\
                                    &&& Omni & Sep-16-2022 & 4.2 & Ethereum PoW & BB \\
                                    \cmidrule{2-8}
                                    
                & V9. Fake event emission & A2(\CS) and A3(\BOS) & THORChain & Jun-29-2021 & 0.35 & THORChain & Au \\
                    &&& THORChain & Jul-16-2021 & 8 & THORChain & Au \\
                    &&& THORChain & Jul-23-2021 & 8 & THORChain & Au \\
                    &&& pNetwork & Sep-19-2021 & 13 & Bnb & UVC;BB \\
                    &&& Cennznet & May-08-2022 & 0.4 & Ethereum & UVC;EOS \\
        \cmidrule{1-8}

        FI & V10. Front-end phishing & A1(\BOS) & EVODeFi &  Mar-22-2022 & 0.3 & Bnb & BB \\
                &&& Celer cBridge & Aug-17-2022 & 0.2 & Ethereum & BB \\ 
      \bottomrule
    \end{tabular}}
    
    \caption{\small A taxonomy of cross-chain bridge attacks and vulnerabilities based on reported attacks (from July 2021 to July 2023). 
    {\bf Type:} PI: Permission Issue; LI: Logic Issue; EI: Event Issue; OI: Other Issue; FI: Front-end Issue.
    {\bf Attack Surface:} $A_{i}$ is the index of possible attacks defined in~\secref{sec:vuls}; \CS, \CD, and \BOS are involved identities in~\figref{figs:workflow}.
    {\bf Defense} (presented in~\secref{sec:defense}): Au: Auditing; BB: Bug Bounty; HF: Hard Fork of Blockchain; UVC: Upgrading Vulnerable Contract; RIA: Revoking Infinite Approval; IVD: Increasing Validator Decentralization; EOS: Examining Off-chain Server.}
    \label{tbl:attack-class}
    \vspace{-13pt}
\end{table*}

\subsection{Permission Issue}
Permission issues are quite common in both regular smart contracts and cross-chain bridges; 
17 out of the total 31 attacks fall into this category. 
We divide this category into five types, including unchecked intermediary permission, misused proof permission, problematic approval permission, invalid signature permission, and leaked key permission.

\subsubsection{V1. Unchecked Intermediary Permission} 
This type of attacks can happen in either \CS or \CD, as shown in~\figref{figs:workflow}.
A smart contract can be assigned a set of managers for permission check when some fundamental changes are made, such as owner modification or money transfer. If the functions in the smart contract are restricted with these permission checks, they can only be called when the caller belongs to one of the managers. 
However, if the caller utilizes the manager as the intermediate contract to call the permissioned functions, the permission check will be bypassed and the attack will happen.
Moreover, we illustrate this type with 2 examples of Solidity code and 1 real-world attack.
\looseness=-1

\begin{figure}[!htbp]
\setlength{\fboxsep}{0pt}%
\setlength{\fboxrule}{0pt}%
\centering
\begin{lstlisting}[language=Solidity]
contract A {
  address owner;
  address manager;
  
  modifier onlyManager {
    require(msg.sender == manager);
  }
  
  function changeOwner(address account) public onlyManager {
    owner = account;
  }
}

contract B {
  function call(address callee, bytes4 funcSig,  bytes parameters) public {
    callee.call(funcSig, parameters);
  }
}
\end{lstlisting}
\caption{Unchecked intermediary permission.}
\label{figs:case1}
\vspace{-10pt}
\end{figure}

\bheading{Example 1.}
We illustrate the pattern of this type
via the example in ~\figref{figs:case1}, in which the smart contract B is the manager of smart contract A. We explain some primitives needed in this example, as follows.
\begin{packeditemize}
    \item \textit{address} is a type within Solidity smart contracts, representing an Ethereum account. 
    \item \textit{bytes4} and \textit{bytes} are both array of bytes, and \textit{bytes4} is a fixed-length array of four bytes.
    \item \textit{modifier} restricts the behaviors of the functions, and can be considered as a compile-time code roll-up.
    \item \textit{function} is a combination of lines of code. In particular, there are four types of functions: external, internal, public, and private. External functions can be called from outside smart contracts and internal functions can only be called internally within the current contract. Moreover, public functions can be called without restrictions and private functions are only visible in the contract they are defined. 
    \item \textit{call} is a default low-level function for interacting with public and external smart contract functions. 
\end{packeditemize}

In this example, the function \textit{changeOwner()} is to modify the owner of the smart contract A to be the parameter \textit{account}. However, not anyone is allowed to call this function, unless the caller \textit{msg.sender} is the manager of the contract, as defined by the modifier \textit{onlyManager}. 
In smart contract B, the function \textit{call()} interacts with any callable function in smart contract \textit{callee}, as long as the correct function signature~\textit{funcSig}, parameters~\textit{parameters}, and the contract address~\textit{callee} are provided. 
If the data of function \textit{call()} in contract B is constructed correctly, anyone can bypass the permission check in contract A, to call \textit{changeOwner()} and transfer the ownership to its account, since contract B is the manager of contract A. 

\bheading{Example 2.}
An alternative vulnerable contract A is that there is a permissioned token transfer function with the modifier \textit{onlyManager}, similar to the function \textit{changeOwner()}. In this situation, the money in contract A will be stolen directly by the attacker.

\bheading{Real-world attacks.} Among the 31 real-world attacks, one falls under this specific vulnerability type.

\iheading{Poly Network attack.}
On Aug-10-2021, Poly Network~\cite{polynetwork} suffered from an attack resulting in \$600M loss. 
In particular, the bridge failed to set the appropriate permission in the manager smart contract and allowed anyone to call the key function in the contract controlled by the manager~\cite{polynetwork-attack}. Thus, the attacker could change the owner to be his/her account and drained all the tokens from the hacked smart contract. 

\subsubsection{V2. Misused Proof Permission} 
This type of attack can occur in step \ding{188} inside \CS or \CD, as shown in~\figref{figs:workflow}. 
In cross-chain bridges, unique proofs should be provided to prove the deposit or burn actions launched by users, for future minting or releasing the related tokens to users. If the smart contract is not designed carefully to deal with proofs, the previous valid proof could be replayed multiple times. As a result, the money will be stolen. We also illustrate this type with one code example and two real-world attacks.

\begin{figure}[!htbp]
\setlength{\fboxsep}{0pt}%
\setlength{\fboxrule}{0pt}%
\centering
\begin{lstlisting}[language=Solidity]
contract A {
  function verifyProof(bytes data) external {
    if isValid(data) return throw;
    receiver, tokenAddr, tokenAmount = decode(data);
    tokenAddr.transfer(receiver, tokenAmount);
  }
}
\end{lstlisting}
\caption{Misused proof permission.}
\label{figs:case2}
\vspace{-5pt}
\end{figure}
\bheading{Example 1.}
We illustrate this type with one example in~\figref{figs:case2}. In particular, if the proof \textit{data} is verified to be valid via the function \textit{isValid()}, the money will be sent back to the user according to the decoded parameters via the function \textit{decode()}. However, the \textit{data} could be constructed maliciously multiple times from one valid proof related data, if the function \textit{isValid()} does not check the uniqueness of the value \textit{data} correctly.

\bheading{Real-world attacks.} Among the 31 real-world attacks, two fall under this specific vulnerability type.

\iheading{Plasma attack.}
Plasma~\cite{plasma} is one of the popular bridges on Polygon, supporting token deposit and withdraw. 
On Oct-05-2021, a vulnerability was found and exploited by white hats rescuing around \$850 million~\cite{plasma-attack}. In this exploit, the previous valid proof could be replayed multiple times, to exit the burn transaction and steal tokens. Fortunately, no loss was caused due to the rescue. 

\iheading{Binance Bridge attack.}
On Oct-06-2022, attackers stole around \$566 million from Binance Bridge~\cite{bnb-attack}. The root cause is a smart contract bug that allows attackers to forge proofs to pass the validation so that the fraudulent withdraw can succeed. To mitigate this attack, Binance-backed blockchains launched a hard fork on Oct-12-2022~\cite{bnb-hard}. 

\subsubsection{V3. Problematic Approval Permission}
This type of attacks can happen in \CS or \CD, as shown in~\figref{figs:workflow}.
To save on transaction gas fees, users must give their permissions to DApps in advance, allowing bridges to transfer their tokens. Such behaviors are
usually implemented by the standard ERC20 function \textit{approval()}, which is insecure when the DApp smart contract cannot handle the permission check correctly. Moreover, we illustrate this type with two examples of Solidity code and three real-world attacks.
\looseness=-1

\begin{figure}[h]
\begin{subfigure}[b]{1\columnwidth}
\setlength{\fboxsep}{0pt}%
\setlength{\fboxrule}{0pt}%
\centering
\begin{lstlisting}[language=Solidity]
contract A {
  function transferWithPermit(address from, address to, address token, uint amount, uint deadline, uint8 v, bytes32 r, bytes32 s) external {
    address _underlying = DERC20(token).underlying();
    IERC20(_underlying).permit(from, address(this), amount, deadline, v, r, s);
    _underlying.call(abi.encodeWithSelector(0x23b872dd, from, token, amount));
  }
}
\end{lstlisting}
\caption{Example 1}    
\label{figs:case3}
\end{subfigure}
\vfill%
\begin{subfigure}[b]{1\columnwidth}
\begin{lstlisting}[language=Solidity]
contract A {
  function transfer(address from, address to, address token, uint amount) public {
    token.call(abi.encodeWithSelector(0x23b872dd, from, to, amount));
  }
}
\end{lstlisting}
\caption{Example 2}    
\label{figs:case4}
\end{subfigure} 
\caption{Problematic approval permission.}
\label{figs:insecure}
\vspace{-10pt}
\end{figure} 

\bheading{Example 1.}
We illustrate this type with the example from~\figref{figs:case3}, in which the function \textit{transferWithPermit()} is not handled correctly. Here, we assume some users have given infinite approvals to the DApp in advance. The vulnerability can be exploited via the following steps.
(1) Calling to an unverified smart contract. In line 3, the attacker can set the address \textit{token} to a malicious contract, which is usually unknown and unverified. Then the attacker can control the returned value, which is not checked by contract A. 
(2) Entering a fallback function. In line 4, the function \textit{permit()} intends to ensure the ability of the current contract with the supplied transaction (v,r,s), which allows transferring users' tokens. However, \textit{permit()} does not exist in the attacker-controlled malicious contract, and the fallback function will be called. If the fallback function is empty and the transaction continues to execute, permit checking will be bypassed. 
(3) Stealing money. Tokens will be transferred from the victim user to the attacker in this step. In line 5, the transfer is made via the low-level call function by constructing the appropriate malicious data, specifying the standard function signature (0x23b872dd) and parameters of \textit{transferFrom()}. 

\bheading{Example 2.}
Another example is shown in~\figref{figs:case4}, in which the function \textit{transfer()} can be called by anyone and then attacks occur, given that the victim (the \textit{from} address in the function parameters) has already approved the token transfer.

\bheading{Real-world attacks.} Among the 31 real-world attacks, three fall under this specific vulnerability type.

\iheading{Multichain attack.} 
Multichain~\cite{multichain} is a cross-chain bridge that supports financial services on multiple blockchains.
On Jan-18-2022, Multichain was attacked, in which several blockchains were affected~\cite{multichain-attack}. It caused around \$3 million loss by multiple group of attackers.
In particular, users approved the infinite number of tokens to bridge-related smart contracts in advance. The attacker exploited the incorrect permission check to pass the validation and drain the tokens from users. 

\iheading{LI.FI attack.}
On May-20-2022, an attacker exploited the vulnerable swapping feature of LI.FI bridge~\cite{lifi}’s smart contract~\cite{lifi-attack}. Instead of swapping, attackers steal money from users by directly transferring their tokens. As a result, anyone who gave infinite approval to the vulnerable contract could be attacked. 

\iheading{Rubic attack.}
On Dec-25-2022, Rubic suffered a hack which resulted in a loss of \$1.4 million~\cite{rubic-attack2}. The main reason behind the attack was the incorrect addition of USDC tokens to the whitelist. This led to the theft of USDC tokens from users to the Rubic contract.

\subsubsection{V4. Invalid Signature Permission} This type of attack can occur at step \ding{184} inside \CD, as shown in~\figref{figs:workflow}. In cross-chain bridges, signatures of validators are required to sign transactions, in order to transfer tokens back to users. In particular, the signatures should be checked. If signature verification is not properly processed, attacks can occur. Moreover, we illustrate this type with 1 example of Solidity code and 3 real-world attacks.

\begin{figure}[!htbp]
\setlength{\fboxsep}{0pt}%
\setlength{\fboxrule}{0pt}%
\centering
\begin{lstlisting}[language=Solidity]
contract A {
  function mintToken(bytes[] signatures, address to, address token, uint amount) public {
    bool ret = verify_signatures(signatures);
    require(ret, "Signatures are not valid!");
    mint(to, token, amount);
  }
}
\end{lstlisting}
\caption{Invalid signature permission.}
\vspace{-5pt}
\label{figs:case5}

\end{figure}
\bheading{Example 1.}
From the example in ~\figref{figs:case5}, \textit{verify\_signatures()} function in line 3 incorrectly checks signatures, and the permission check is bypassed by the attacker. Then, the attacker can mint tokens without any cost since he/she has the permission due to the forged but passed signatures.

\bheading{Real-world attacks.} Among the 31 real-world attacks, three fall under this specific vulnerability type.

\iheading{ChainSwap attack.}
ChainSwap~\cite{chainswap} is a cross-chain asset bridge, which allows users to bridge tokens between blockchains seamlessly. ChainSwap was exploited on Jul-10-2021 and many projects using ChainSwap were affected, including Razor Network~\cite{razor}. The attacker stole tokens worth \$8 million and sold the tokens on various exchanges~\cite{chainswap-attack}. 

\iheading{Wormhole attack.}
On Feb-02-2022, an attacker launched this attack with bypassing the verification process in the Wormhole bridge~\cite{wormhole} and minted Wormhole \textit{ETH} (\textit{wETH}) tokens~\cite{wormhole-attack}. In particular, the attacker injected a faked account and successfully generated a malicious message to mint 120,000 \textit{wETH} worth \$320 million. 

\iheading{Multichain attack.}
On Feb-15-2022, multichain attack took place again by the insufficient signature verification~\cite{multichain-attack2}. The attacker managed to transfer the approved tokens to the victim's contract, resulting in a profit of 87 ETH.

\subsubsection{V5. Leaked Key Permission} This type of attack can occur in \BOS, as shown in~\figref{figs:workflow}.
Attackers usually compromise the private key of admin bridge accounts or enough private keys of bridge validators. We illustrate eight real-world attacks caused by such vulnerability.

\bheading{Real-world attacks.} Among the 31 real-world attacks, eight fall under this specific vulnerability type.

\iheading{WonderHero attack.}
On Apr-07-2022, WonderHero, a NFT-based game, claimed that it was attacked by attackers who obtained signatures of validators and then minted 80 million \textit{WND} tokens~\cite{wonderhero-attack}. The loss was approximately \$0.3 million. 

\iheading{Ronin Network attack.} 
Ronin Network~\cite{ronin} suffered from the attack on May-29-2022, which was caused by the compromised private keys~\cite{ronin-attack}. In particular, the attacker accessed four validators controlled by Sky Mavis~\cite{sky-mavis} and the fifth validator from Axie DAO~\cite{axie-dao}. Due to the 5-of-9 multi-signature mechanism, the attack succeeded and caused a \$625 million loss. 

\iheading{Horizon Bridge attack.}
Horizon Bridge~\cite{harmony} supports connections between any Proof-of-Work (PoW) and Proof-of-Stake (PoS) blockchain. On Jun-24-2022, the bridge was hacked and lost around \$100 million. The attackers compromised private keys of two validators and successfully stole tokens with the 2-of-4 multi-signature mechanism~\cite{harmony-attack}.

\iheading{QAN platform bridge attack.}
On Oct-11-2022, the QAN platform bridge experienced a security breach that was caused by the compromise of a private key~\cite{qan-attack}. The attacker gained access to the private key of the bridge address and withdrew tokens from both the Ethereum and BNB blockchains. The attack resulted in a total loss of approximately \$2 million at the time it occurred.

\iheading{Rubic attack.}
Attackers compromised the private key of an admin wallet address belonging to Rubic on Nov-02-2022~\cite{rubic-attack}. This wallet was responsible for managing the RBC/BRBC bridge and staking rewards. The attackers were able to transfer 34 million RBC out of the wallet, which was valued at \$1.2 million at the time. 

\iheading{pNetwork attack.}
On Nov-04-2022, more than \$1 billion of pGALA tokens on Bnb were minted out of thin air. This occurred because the private key was leaked on Github~\cite{pnetwork-attack2}. To prevent further attacks, pNetwork minted a substantial number of tokens themselves, trying to the drainage of the liquidity pool. However, this action was not handled correctly and introduced losses caused by arbitrage opportunities.

\iheading{Poly Network attack.} 
On Jul-01-2023, attackers were suspected to take control of the private keys associated with Poly Network wallets, granting themselves the ability to create signatures without limit. This exploit allowed the hacker to generate 57 tokens that span 10 different blockchains, including Ethereum and BNB Chain, among others. As a result of this attack, a total of 5,196.95 ETH, equivalent to a staggering \$10,201,612 USD in losses, was incurred~\cite{polynetwork-attack2}.

\iheading{Multichain attack.} On Jul-06-2023, Multichain experienced a significant breach, resulting in substantial cross-chain fund losses (\$231,129,033). Funds were consolidated into a single EOA across 9 chains, suggesting full control by the attacker and raising internal operation suspicions~\cite{polynetwork-attack2}.

\subsection{Logic Issue}
Logic issue are caused by the logical mistakes in smart contracts. We classify this type into incorrect balance logic and inaccurate initialization logic. Four of the total 31 attacks belong to this category. 

\subsubsection{V6. Incorrect Balance Logic}
This type of attack can occur in either \CS or \CD, as shown in~\figref{figs:workflow}.
A liquidity pool is a collection of locked tokens. 
Assume there are two types of tokens in a liquidity pool, the quantities of those assets are called reserves. 
Generally, $x \times y = c$ ($x$ and $y$ are the reserves of the two types of tokens; c is a constant number). 
However, the value paid by the user should be slightly higher than the expected amount every time the tokens are deposited or withdrawn to pay the fee. Thus, to ensure the correctness of such mechanisms, the balances of these two tokens should be no less than the initial constant. Otherwise, any user can steal money from the pool. Moreover, we illustrate this type with 1 example of Solidity code and 2 real-world attacks.

\begin{figure}[!htbp]
\setlength{\fboxsep}{0pt}%
\setlength{\fboxrule}{0pt}%
\centering
\begin{lstlisting}[language=Solidity]
contract A {
  function balanceCheck() public {
    _reserve0, _reserve1 = getReserves();
    uint balance0Adjusted = balance0.mul(1000).sub(amount0In.mul(3));
    uint balance1Adjusted = balance1.mul(1000).sub(amount1In.mul(3));
    require(balance0Adjusted.mul(balance1Adjusted) >= uint(_reserve0).mul(_reserve1).mul(1000**2), 'Balance check failed!');
  }
}
\end{lstlisting}
\caption{Incorrect balance logic.}
\label{figs:case6}
\vspace{-5pt}
\end{figure}

\bheading{Example 1.}
An example is shown in~\figref{figs:case6}, which is the representative source code of the balance check for a liquidity pool. The vulnerability exists if line 6 is removed.

\bheading{Real-world attacks.} Among the 31 real-world attacks, two fall under this specific vulnerability type.

\iheading{PolkaBridge attack.}
PolkaBridge~\cite{polkabridge} was attacked on Nov-22-2021 and lost around \$0.6 million. The attack was caused by a smart contract bug after the balance check was removed mistakenly.

\iheading{Poolz Finance attack.} Poolz Finance was attacked on May-15-2023, in which attackers triggered an integer overflow vulnerability and stole around \$665,000 on several blockchains~\cite{poolz-finance-attack}.

\subsubsection{V7. Inaccurate Initialization Logic} 
This type of attack can occur in \CS or \CD, as shown in~\figref{figs:workflow}.
The initialization process sets some privileged addresses as managers or owners of the smart contracts. Similarly, some non-trivial variables (e.g., verification variable) will be assigned with initial values. If these fundamental variables are not defined appropriately, bridges will be attacked. Moreover, we illustrate this type with a code example and two real-world attacks.

\begin{figure}[!htbp]
\setlength{\fboxsep}{0pt}%
\setlength{\fboxrule}{0pt}%
\centering
\begin{lstlisting}[language=Solidity]
contract A {
  address initializer_addr;
  address owner;
  bool canTransfer;
  
  modifier initializer {
    msg.sender = initializer_addr;
  }
  
  function initialize() public initializer {
    owner = 0x; 
    // or
    canTransfer = true;
  }
}
\end{lstlisting}
\caption{Inaccurate initialization logic.}
\label{figs:case7}
\vspace{-10pt}
\end{figure}

\bheading{Example 1.}
~\figref{figs:case7} gives an example in which line 11 sets the owner of the contract as an incorrect address, or line 13 defines the key variable with incorrect value.

\bheading{Real-world attacks.} Among the 31 real-world attacks, two belong to this specific vulnerability type.

\iheading{Nomad attack.}
Nomad~\cite{nomad} suffered from an attack on Aug-02-2022 with around \$190 Million loss. 
For example~\cite{nomad-attack-certik}, an attacker~\cite{0xa8c83b1b} sent 0.01 \textit{WBTC} to Nomad in Moonbeam transaction~\cite{0xcca9299c}
and then Nomad transferred 100 \textit{WBTC} back to the attacker
in Ethereum transaction~\cite{0xa5fe9d04}
This attack was caused by an initialization error, in which the trusted root was set to an incorrect value, the permission check could be passed by anyone, and the money was then stolen by attackers~\cite{nomad-attack}. 

\iheading{Allbridge attack.} 
On Apr-01-2023, Allbridge fell victim to a flash loan attack triggered by an error in the price calculation logic. This exploit resulted in approximately \$570,000 worth of Bnb being lost~\cite{polynetwork-attack2}.

\subsection{Event Issue}
To gather the necessary information from the transactions, bridges usually fetch events from smart contracts and interpret these events to perform future steps. For example, if some tokens are sent from users to bridges, the related \textit{Deposit} events will be emitted, serving as the deposit proof. However, events can be problematic and money will be stolen from bridges. In particular, this type includes incorrect event emission and fake event emission. Moreover, 8 of the total 31 attacks belong to this category. 

\subsubsection{V8. Incorrect Event Emission}
This type of attacks can happen in step \ding{182} inside \CS, as shown in~\figref{figs:workflow}. 
These two functions \textit{deposit()} and \textit{depositETH()} emit the same \textit{Deposit} event, representing that users deposit \textit{ETH} tokens to the bridge. Then, \textit{BOS} will process the events, minting \textit{ETH} tokens to users no matter what tokens are deposited. Attackers could utilize this vulnerability to deposit less valuable tokens (e.g., USDT~\cite{usdt}) and earn \textit{ETH} tokens back. Moreover, we illustrate this type with two examples of Solidity code and three real-world attacks.

\begin{figure}[h]
   \begin{subfigure}[b]{1\columnwidth}
           \setlength{\fboxsep}{0pt}%
\setlength{\fboxrule}{0pt}%
\centering
        \begin{lstlisting}[language=Solidity]
contract A {
  function deposit(address from, address to, address token, uint amount) external payable {
    token.transferFrom(from, to, amount)
    emit Deposit(from, to, amount);
  }

  function depositETH() external payable {
    transferETH(msg.sender,address(this),msg.value);
    emit Deposit(msg.sender,address(this),msg.value);
  }
}
\end{lstlisting}
        \caption{Example 1}    
        \label{figs:case8}
    \end{subfigure}
  \vfill%
   \begin{subfigure}[b]{1\columnwidth}
\begin{lstlisting}[language=Solidity]
contract A {
  function deposit(address from, address to, address token, uint amount) external payable {
    if (token != wrapped_token) {
      lockOrBurnToken(from, to, token, amount);
    }
    emit Deposit(from, to, amount);
  }
        
  function depositETH() external payable {
    transferETH(msg.sender,address(this),msg.value);
    emit Deposit(msg.sender,address(this),msg.value);
  }
}
\end{lstlisting}
        \caption{Example 2}    
        \label{figs:case9}
    \end{subfigure} 
  \caption{Incorrect event emission.}
  \vspace{-10pt}
  \label{figs:confused}
  
\end{figure} 
\bheading{Example 1.}
In the example shown in~\figref{figs:case8}, the attacker transferred some less valuable tokens (e.g., USDT~\cite{usdt}) via the \textit{deposit()} function. The attacker can get wrapped \textit{ETH} tokens. Moreover, the attacker can even construct malicious data to transfer nothing, but obtain a \textit{Deposit} event. For instance, the token address can be some invalid addresses and the transaction will not fail, since the return value of function \textit{transferFrom()} is not checked.
Then, the attacker can use the event to steal tokens.

\bheading{Example 2.}
Another example shown in~\figref{figs:case9} has the same problem that \textit{Deposit} events cannot be distinguished from \textit{deposit()} and \textit{depositETH()}. The difference is that, in this example, a lock or burn action on the transferred tokens is performed when the tokens are not wrapped token. However, if the attacker assigns the tokens to be wrapped tokens, the permission check will be bypassed and the \textit{Deposit} events will be emitted incorrectly.

\bheading{Real-world attacks.} Among the 31 real-world attacks, three fall under this specific vulnerability type.

\iheading{Qubit attack.}
Qubit~\cite{qubit} exploit happened on Jan-27-2022, in which the attacker used the vulnerable functions and obtained an ETH \textit{Deposit} event with a ERC20 token transfer. The incorrect event allowed the attacker to mint around 77,162 \textit{qxETH}, which was worth around \$80 million.  

\iheading{Meter.io attack.}
On Feb-05-2022, the Meter.io bridge~\cite{meter} project was exploited, causing a \$4.2 million loss~\cite{meter-attack}. In the attack, the attacker obtained \textit{Deposit} events by using the function \textit{deposit()} targeting \textit{WETH}/\textit{WBNB} tokens without real deposit. 

\iheading{Omni bridge replay attack.} On Sep-16-2022, the Omni Bridge replay attack was due to a chainID permission check vulnerability~\cite{omni-attack}. Replay attacks can occur during blockchain protocol upgrades or when a vulnerability is exploited across multiple blockchains. The attacker in this case first transfer 200 \textit{WETH} to Ethereum Proof-of-Stake (PoS) blockchain and then used an outdated chainID to replay the same transaction on the Ethereum Proof-of-Work (PoW) blockchain~\footnote{There are users who did not switch to PoS and are still on the Ethereum PoW blockchain.}.

\subsubsection{V9. Fake Event Emission} 
This type of attack can only occur when both \CS and \BOS have bugs in the first two steps (\ding{182} and \ding{183}), as shown in~\figref{figs:workflow}.
In particular, the \textit{Deposit} event is being emitted from the attacker controlled smart contract, not from the bridge-related contract. However, the bridge off-chain server is not aware of the problem and mistakenly accepts fake events. Thus, the attackers can deposit nothing, fake the event, and steal tokens.
Moreover, we illustrate this type with one Solidity code example and five real-world attacks.

\begin{figure}[!htbp]
\setlength{\fboxsep}{0pt}%
\setlength{\fboxrule}{0pt}%
\centering
\begin{lstlisting}[language=Solidity]
contract Bridge {
  function withdraw(address recipient, uint256 amount) external payable {
    bool sent, _ = recipient.call{value:amount}("");
    require(sent, "Sent failed!");
  }
}

contract Attacker {
  function() external payable {
    emit Deposit(msg.sender,address(this),msg.value);
  }
}
\end{lstlisting}
\caption{Fake event emission.}
\vspace{-5pt}
\label{figs:case10}

\end{figure}
\bheading{Example 1.}
An example is shown in ~\figref{figs:case10}. If the attacker withdraws money from the contract Bridge, the fallback function in the appointed smart contract \textit{recipient} will be called. Then the attacker can fake the \textit{Deposit} event in the fallback function within the Attacker contract, to steal tokens.

\bheading{Real-world attacks.} Among the 31 real-world attacks, five fall under this specific vulnerability type.

\iheading{THORChain attack.} In a series of unfortunate events, THORChain experienced three separate fake deposit attacks in quick succession. The first attack occurred on Jun-29-2021, causing a loss of approximately \$350,000. Following closely, a second attack struck on Jul-16-2021, resulting in a staggering loss of nearly \$8 million. Just a week later, on Jul-23-2021, THORChain was targeted yet again, resulting in another substantial loss of nearly \$8 million. These incidents underscored the vulnerability of the THORChain network to such fraudulent deposit exploits during this period~\cite{THORChain-attack}.

\iheading{pNetwork attack.}
On Sep-19-2021, the attacker generated counterfeit events to access cryptocurrency in the Bitcoin blockchain~\cite{pnetwork-attack}. The events were executed without confirmation of their legitimacy from the pNetwork contracts. Thus, the attacker successfully stole 277 \textit{BTC}, equivalent to more than \$13 million by falsifying the events.

\iheading{Cennznet attack.}
On May-08-2022, an attacker hacked the Cennznet bridge~\cite{cennznet} by emitting fake \textit{Deposit} events. The bridge captured the fake events and sent ETHs to the attacker. In total, \$0.4 million (around 155 \textit{ETH}) was stolen.

\subsection{Front-end Issue}
Some bridges suffer from front-end hijackings, in which websites or smart contracts are redirected to fake websites or contracts created by attackers. Specifically, two of the total 31 attacks belong to this category.

\subsubsection{V10. Front-end Phishing} 
Front-end phishing refers to a type of cyberattack where the attacker manipulates the user interface or the frontend of the bridge to deceive users into revealing sensitive information or taking malicious actions unwittingly. We illustrate this type with two real-world attacks.

\bheading{Real-world attacks.} Among the 31 real-world attacks, two of them belong to this specific type of vulnerability.

\iheading{EVODeFi attack.}
On Mar-22-2022, an attacker gained unauthorized access to the EVODeFi domain by exploiting a bug in Vercel, which hosts the bridge's front-end~\cite{evodefi-attack}. The attacker was able to bypass the domain transfer confirmation process, which should have been required from the legitimate owner, EVODeFi. As a result of the security breach, the attacker was able to steal \$320,000.

\iheading{Celer cBridge attack.} 
On Aug-17-2022, Celer cBridge suffered from a domain name system (DNS) poisoning attack, resulting in a loss of \$240,000~\cite{celer-attack}. Some users were redirected to malicious smart contracts while using the bridge, and their approved tokens were drained by the attacker.

\section{Defenses and Recommendations}
\label{sec:defense}

\subsection{Existing Defenses}
We summarize the existing defenses from real-world attacks, categorize the defenses into four types: general defenses and defenses of three types of attacks, including PI, LI, EI, FI (defined in~\secref{sec:taxonomy}), and illustrate the defenses with examples. 


\subsubsection{General Defenses} 
The general defenses include auditing, bug bounty, and upgrading vulnerable contracts, which can be applied to any attack. 

\iheading{Auditing (Au).} DApps typically are audited by trusted third parties (e.g., Certik~\cite{certik}) before being deployed. The more comprehensively a DApp is audited, the more secure the project will be. \looseness=-1

\iheading{Bug Bounty (BB).} Bug bounty programs can not only motivate white hats to disclose vulnerabilities before being attacked, but also serve as rewards for attackers to send the stolen funds back. For example, Wormhole offered a \$10m bug bounty reward in exchange for lost funds from attackers~\cite{wormhole-after}.

\iheading{Hard Fork (HF).} To mitigate severe attacks (e.g., DAO attack on Ethereum), the blockchain can be upgraded by a hard fork that requires all nodes and users to update. 

\iheading{Upgrading Vulnerable Contracts (UVC).} DApps can fix smart contract bugs by upgrading contracts. For example, Multichain removed the vulnerabilities in the token liquidity pool and upgraded new contracts~\cite{multichain-after}. 

\subsubsection{Defenses of PI Attacks} Permission issue (PI) attacks are caused by incorrect permission checks or problematic protocols (e.g., multiple signatures).

\iheading{Revoking Infinite Approval (RIA).} Developers should avoid asking for unlimited user approvals, as it can be dangerous. Instead, they should request only the necessary approvals. For example, Multichain advised its users to revoke approvals on the same day of the attack (Jan-18-2022)~\cite{multichain-after2}.

\iheading{Increasing Validator Decentralization (IVD).} To mitigate the vulnerability of key permissions that have leaked, bridges can improve their security by increasing the number of validators or improving their decentralization. For example, Ronin Network increased the number of validators to 21 after being attacked~\cite{ronin-after}. 

\subsubsection{Defenses of LI Attacks} 
Smart contract vulnerabilities stemming from logical errors can lead to logic issue attacks. For example, PolkaBridge~\cite{polkabridge} attack happened due to the incorrect calculation of the token price. Bridges typically employ standard defensive measures to combat such attacks.  

\subsubsection{Defenses of EI Attacks}
Bridges off-chain server monitors on-chain activities by recording events and performing related actions. If such events are incorrect, event issue (EI) attacks can occur.  

\iheading{Examining Off-chain Server (EOS).} 
Bridges extract information from on-chain transactions, 
store them in off-chain datasets, 
and perform operations
such as submitting transactions to transfer tokens back to users.
It is common to have vulnerabilities on the off-chain server
since they are not public and may not be audited by others.
Thus, it is non-trivial to examine the implementation of off-chain server closely and cooperate with the trusted third parities for auditing, to avoid risks. 

\subsubsection{Defenses of FI Attacks} 
Front-end issue (FI) attacks usually arise when malevolent websites redirect user transfers to attackers. Bridges employ standard defensive strategies, such as bug bounties, to prevent or mitigate such attacks.

\subsection{Recommendations to Developers}
Based on our study, this section offers recommendations for developers building on-chain backends comprising smart contracts (for blockchain transactions) and off-chain servers (for off-chain operations).

\subsubsection{On-chain Backend.} For on-chain backend security, developers should pay particular attention to smart contracts.

\iheading{Ensuring proper permission handling code.} Developers must thoroughly review all possible execution paths regarding permission checks. Additionally, they can use tools such as Manticore~\cite{mossberg2019manticore} to assess the correctness of the code.

\iheading{Avoiding unverified smart contract calls.} To prevent attacks where attackers deploy malicious smart contracts and hijack the control flow to their code, developers should avoid calling unverified smart contracts, particularly user-specified contracts, unless absolutely necessary.

\iheading{Checking return values of external calls.} Developers must be mindful of return values of functions, especially external function calls, as attackers can leverage inconsistencies caused by unchecked return values to launch exploits.

\iheading{Avoiding directly copying code.} While copying smart contract code from popular projects may save time, it can also result in code with bugs or vulnerabilities. Therefore, developers should closely examine the code and understand its logic.

\iheading{Setting longer withdraw latency.} Increasing the latency on token withdrawal can provide developers with additional time to mitigate risks. However, it may also affect bridge performance, and developers must weigh the trade-off between security and speed.

\iheading{Providing insurance.} Bridges can establish insurance for assets to compensate users in case of an attack, similar to real-world financial services. While precautions can reduce attack risks, insurance can mitigate users' financial losses in the event of an attack.

\subsubsection{Off-chain Server.} For off-chain server security, developers can monitor bridge server activities.

\iheading{Money monitoring.} 
DApps must vigilantly monitor their transactions to prevent losses from copycat or 1-day attacks. For instance, Ronin~\cite{ronin-attack} was aware of the attack several days after it occurred, avoiding further attacks. DApps must keep a close eye on transactions with unusual behaviors, such as large transfers or the minting and burning of tokens. It is recommended to receive early alerts for such transactions.

\section{Open Problems and Future Directions}
\label{sec:question}


\subsection{Security Property Concerns}

\bheading{Q1: Future of bridges: will it be multi-chain?} With so many severe attacks on bridges, it raises doubts on the security of cross-chain bridges. There 
are suggestions of replacing existing cross-chain bridges with the multi-chain solutions, 
which are blockchains of blockchains (e.g., Cosmos). 
According to Vitalik Buterin, the founder of Ethereum, 
multi-chain is likely to dominate the future considering it is more secure~\cite{multichain}. 
To this end,
a systematic study measuring and comparing the cross-chain bridges and multi-chain protocols should be performed. 

\bheading{Q2: Trust models: what is the real security level of existing bridges?} 
In the official documentations of bridges with trusted or semi-trustless models (e.g., Horizon Bridge~\cite{horizon-bridge}), 
it is always stated that they are secured with the guardian from trusted third parties. 
It is unclear what the \textit{true} security level of these bridges is. 
For example, bridges with individual validators are more secure than bridges with validators controlled by a single party. 
Moreover, trustless bridges claim to have the same security as the underlying public blockchain. 
We should be aware of the \textit{true} security level of these bridges. 
To answer this question, systematic work studying the security of real-world cross-chain bridges with different trust levels is needed. 


\subsection{Bridge Related Issues}




\bheading{Q3: Bridge usages: any financial crime issues?} 
Cross-chain bridges are now often used for cross-chain transfers and swaps.
It is not clear whether these bridges are used for any illegal purposes, such as money laundering.
To avoid being traced and deanonymized, 
attackers usually mix their stolen assets;
the assets will be exchanged between different users, making it difficult to trace their original accounts. 
In particular, many attacks move the stolen funds to bridges in the end. 
For example, attackers in Hamster rugpull profited 1,730
\textit{BNB} tokens along with the Cheese rugpull~\cite{hamster}, in which 900 \textit{BNB} tokens were sent to the Wormhole bridge. 
After cross-chain transfer via bridges, 
the behaviors of attackers are hard to analyze and the assets are difficult to trace. 
With the recent sanction of Tornado Cash, which is a popular token mixer~\cite{tornado},
we believe that the use of bridges as cash mixers will become more and more popular.
Thus, work to develop general and systematic frameworks to analyze
illegal activities such as money laundering is definitely in need.

\bheading{Q4: Inconsistencies: are the bridges as efficient and convenient as they claim to be?} 
Except for the security property, 
bridges describe their costs (including both fee and time cost) in their official documentations. 
Are these really the same as what they claim to be? 
For example, if the bridges claim that the cross-chain transactions take no more than 20 minutes, 
can all transactions be completed within that timeframe? 
There might be some inconsistencies between the real-world situations and the ones bridges officially claim. 
To understand the bridges deeper and identify those inconsistencies, bridge-related transactions should be studied. 

\subsection{Attack Detection and Prevention}


\bheading{Q5: Attack detection: How to detect attacks?} 
Many works~\cite{sereum-ndss19,grossman2017online,frank2020ethbmc,torres2018osiris,Zhang2020TXSPECTORUA,jiang2018contractfuzzer,grieco2020echidna} focus on detecting blockchain attacks from the perspectives of smart contracts and transactions. 
However, it is yet unclear how these bridge-related attacks can be detected. 
So far, there has been only one work~\cite{zhang2022xscope} detecting cross-chain bridge attacks; 
Zhang \etal proposes a tool to discover security violations, 
detect bridge attacks from transactions, and evaluate the tool on four cross-chain bridges. 
However, only three attacks presented in~\secref{sec:taxonomy} can be detected by the proposed tool. 
With so many severe attacks,
we believe that there are urgent needs to propose a framework that can detect bridge attacks from transactions or smart contracts. 

\bheading{Q6: Attack prevention: how to prevent attacks?}
Most of existing bridge attacks
are due to smart contract code bugs, and
there will be more attacks in the future. 
Detecting these attacks in a post-mortem manner is not enough, 
since it cannot prevent attacks from happening. 
Solutions to 
prevent these attacks 
or
principled methods to reduce 
attack surfaces
are also important to protect the cross-chain bridges.


\section{Conclusion}
\label{sec:con}

In this paper, we provide the \textit{first} systematic study to understand the security issues of cross-chain bridges.
We first present the models of existing cross-chain bridges,
then identify potential attack surfaces.
We further analyze all 31 bridge attacks over the period of July 2021 to July 2023, 
study the fundamental vulnerabilities that lead to the attacks,
and classify them into three categories based on the vulnerabilities exploited by the attacker.
We summarize existing defenses and provide recommendations for developers to build a secure bridge ecosystem.
We also discuss open problems and future research directions.
Our work provides a holistic view of security issues of cross-chain bridges, 
and the insights are valuable for facilitating future research in the field. 

\newpage

\let\oldbibliography\thebibliography
\renewcommand{\thebibliography}[1]{%
  \oldbibliography{#1}%
  \setlength{\itemsep}{4pt}%
}

{
\scriptsize
\bibliographystyle{unsrt}
\bibliography{references}

\begin{thebibliography}{100}

\bibitem{ethereum}
Ethereum.
\newblock Welcome to ethereum.
\newblock \url{https://ethereum.org/en/}, 2023.

\bibitem{dapp-radar}
DappRadar.
\newblock The world’s dapp store.
\newblock \url{https://dappradar.com/}, 2023.

\bibitem{anyswap}
Anyswap Bridge.
\newblock Anyswap dex user guide.
\newblock \url{https://anyswap-faq.readthedocs.io/en/latest/index.html}, 2023.

\bibitem{solana}
Solana Cashio.
\newblock Solana cashio.
\newblock \url{https://www.solanacash.io}, 2023.

\bibitem{chainbridge}
ChainBridge.
\newblock Chainbridge overview.
\newblock \url{https://chainbridge.chainsafe.io}, 2023.

\bibitem{substrate}
Log Rocket.
\newblock Substrate blockchain development: Core concepts.
\newblock \url{https://blog.logrocket.com/substrate-blockchain-framework-core-concepts/}, 2021.

\bibitem{dezentralizedfinance}
Dezentralized Finance.
\newblock Cross chain bridges.
\newblock \url{https://dezentralizedfinance.com/cross-chain-bridges/}, 2023.

\bibitem{chainswap-attack}
Razor Network.
\newblock Chainswap exploit post-mortem.
\newblock \url{https://medium.com/razor-network/chainswap-exploit-post-mortem-d73f5d15ce3c}, 2021.

\bibitem{nomad-attack}
HALBORN.
\newblock Explained: The nomad hack (august 2022).
\newblock \url{https://halborn.com/explained-the-nomad-hack-august-2022/}, 2022.

\bibitem{chainalysis}
Chainalysis.
\newblock Vulnerabilities in cross-chain bridge protocols emerge as top security risk.
\newblock \url{https://blog.chainalysis.com/reports/cross-chain-bridge-hacks-2022/}, 2022.

\bibitem{zhang2022xscope}
Jiashuo Zhang, Jianbo Gao, Yue Li, Ziming Chen, Zhi Guan, and Zhong Chen.
\newblock Xscope: Hunting for cross-chain bridge attacks.
\newblock {\em arXiv preprint arXiv:2208.07119}, 2022.

\bibitem{buterin2016chain}
Vitalik Buterin.
\newblock Chain interoperability.
\newblock {\em R3 Research Paper}, 9, 2016.

\bibitem{belchior2021survey}
Rafael Belchior, Andr{\'e} Vasconcelos, S{\'e}rgio Guerreiro, and Miguel Correia.
\newblock A survey on blockchain interoperability: Past, present, and future trends.
\newblock {\em ACM Computing Surveys (CSUR)}, 54(8):1--41, 2021.

\bibitem{wang2021sok}
Gang Wang.
\newblock Sok: Exploring blockchains interoperability.
\newblock {\em Cryptology ePrint Archive}, 2021.

\bibitem{zamyatin2021sok}
Alexei Zamyatin, Mustafa Al-Bassam, Dionysis Zindros, Eleftherios Kokoris-Kogias, Pedro Moreno-Sanchez, Aggelos Kiayias, and William~J Knottenbelt.
\newblock Sok: Communication across distributed ledgers.
\newblock In {\em International Conference on Financial Cryptography and Data Security}, pages 3--36. Springer, 2021.

\bibitem{shadab2020cross}
Narges Shadab, Farzin Houshmand, and Mohsen Lesani.
\newblock Cross-chain transactions.
\newblock In {\em 2020 IEEE International Conference on Blockchain and Cryptocurrency (ICBC)}, pages 1--9. IEEE, 2020.

\bibitem{wang2022efficient}
Wenqi Wang, Zhiwei Zhang, Guoren Wang, and Ye~Yuan.
\newblock Efficient cross-chain transaction processing on blockchains.
\newblock {\em Applied Sciences}, 2022.

\bibitem{liang2022privacy}
Xiubo Liang, Yu~Zhao, Junhan Wu, and Keting Yin.
\newblock A privacy protection scheme for cross-chain transactions based on group signature and relay chain.
\newblock {\em International Journal of Digital Crime and Forensics (IJDCF)}, 14(2):1--20, 2022.

\bibitem{su2022cross}
Hong Su, Bing Guo, Jun~Yu Lu, and Xinhua Suo.
\newblock Cross-chain exchange by transaction dependence with conditional transaction method.
\newblock {\em Soft Computing}, 26(3):961--976, 2022.

\bibitem{lee2022sok}
Sung-Shine Lee, Alexandr Murashkin, Martin Derka, and Jan Gorzny.
\newblock Sok: Not quite water under the bridge: Review of cross-chain bridge hacks.
\newblock In {\em 2023 IEEE International Conference on Blockchain and Cryptocurrency (ICBC)}. IEEE, 2023.

\bibitem{chainspot}
Chainspot.
\newblock Chainspot.
\newblock \url{https://chainspot.io/bridge}, 2023.

\bibitem{certik}
Certik.
\newblock Certik blockchain security leaderboard.
\newblock \url{https://www.certik.com/}, 2023.

\bibitem{halborn}
Halborn.
\newblock Halborn blockchain security firm: Ethical hackers, infosec.
\newblock \url{https://halborn.com/}, 2023.

\bibitem{p2p}
Blockchain Council.
\newblock Blockchain \& role of p2p network.
\newblock \url{https://www.blockchain-council.org/blockchain/blockchain-role-of-p2p-network/}, 2023.

\bibitem{sc}
Ethereum.
\newblock Introduction to smart contracts.
\newblock \url{https://ethereum.org/en/developers/docs/smart-contracts/}, 2023.

\bibitem{upgrade-contract}
OpenZeppelin.
\newblock Writing upgradeable contracts - openzeppelin docs.
\newblock \url{https://docs.openzeppelin.com/upgrades-plugins/1.x/}, 2023.

\bibitem{bnb}
BNB.
\newblock Binance smart chain.
\newblock \url{https://github.com/bnb-chain/whitepaper/blob/master/WHITEPAPER.md}, 2023.

\bibitem{fantom}
Fantom Foundation.
\newblock Intro to fantom.
\newblock \url{https://fantom.foundation/intro-to-fantom/}, 2023.

\bibitem{tx}
IBM.
\newblock What is blockchain technology?
\newblock \url{https://www.ibm.com/topics/what-is-blockchain}, 2023.

\bibitem{fee}
Ethereum.
\newblock Gas and fees.
\newblock \url{https://ethereum.org/en/developers/docs/gas/}, 2023.

\bibitem{erc20}
Ethereum.
\newblock Erc-20 token standard.
\newblock \url{https://ethereum.org/en/developers/docs/standards/tokens/erc-20/}, 2023.

\bibitem{erc721}
Ethereum~Improvement Proposals.
\newblock Eip-721: Non-fungible token standard.
\newblock \url{https://eips.ethereum.org/EIPS/eip-721}, 2023.

\bibitem{wtoken}
Coindesk.
\newblock What are wrapped tokens?
\newblock \url{https://www.coindesk.com/learn/what-are-wrapped-tokens/}, 2023.

\bibitem{dapp}
Blockchainhub.
\newblock Decentralized applications – dapps.
\newblock \url{https://blockchainhub.net/decentralized-applications-dapps/}, 2023.

\bibitem{lp}
B2Broker.
\newblock How do liquidity pools work in crypto?
\newblock \url{https://b2broker.com/news/how-do-liquidity-pools-work-in-crypto/}, 2022.

\bibitem{herlihy2018atomic}
Maurice Herlihy.
\newblock Atomic cross-chain swaps.
\newblock In {\em Proceedings of the 2018 ACM symposium on principles of distributed computing}, pages 245--254, 2018.

\bibitem{thyagarajan2022universal}
Sri~AravindaKrishnan Thyagarajan, Giulio Malavolta, and Pedro Moreno-Sanchez.
\newblock Universal atomic swaps: Secure exchange of coins across all blockchains.
\newblock In {\em 2022 IEEE Symposium on Security and Privacy (SP)}, pages 1299--1316. IEEE, 2022.

\bibitem{gugger2020bitcoin}
Jo{\"e}l Gugger.
\newblock Bitcoin-monero cross-chain atomic swap.
\newblock {\em Cryptology ePrint Archive}, 2020.

\bibitem{miraz2019atomic}
Mahdi~H Miraz and David~C Donald.
\newblock Atomic cross-chain swaps: development, trajectory and potential of non-monetary digital token swap facilities.
\newblock {\em arXiv preprint arXiv:1902.04471}, 2019.

\bibitem{borkowski2018towards}
Michael Borkowski, Daniel McDonald, Christoph Ritzer, and Stefan Schulte.
\newblock Towards atomic cross-chain token transfers: State of the art and open questions within tast.
\newblock {\em Distributed Systems Group TU Wien (Technische Universit at Wien), Report}, 8, 2018.

\bibitem{back2014enabling}
Adam Back, Matt Corallo, Luke Dashjr, Mark Friedenbach, Gregory Maxwell, Andrew Miller, Andrew Poelstra, Jorge Tim{\'o}n, and Pieter Wuille.
\newblock Enabling blockchain innovations with pegged sidechains.
\newblock {\em URL: http://www. opensciencereview. com/papers/123/enablingblockchain-innovations-with-pegged-sidechains}, 72:201--224, 2014.

\bibitem{singh2020sidechain}
Amritraj Singh, Kelly Click, Reza~M Parizi, Qi~Zhang, Ali Dehghantanha, and Kim-Kwang~Raymond Choo.
\newblock Sidechain technologies in blockchain networks: An examination and state-of-the-art review.
\newblock {\em Journal of Network and Computer Applications}, 149:102471, 2020.

\bibitem{kiayias2020proof}
Aggelos Kiayias and Dionysis Zindros.
\newblock Proof-of-work sidechains.
\newblock In {\em Financial Cryptography and Data Security: FC 2019 International Workshops, VOTING and WTSC, St. Kitts, St. Kitts and Nevis, February 18--22, 2019, Revised Selected Papers 23}, pages 21--34. Springer, 2020.

\bibitem{gavzi2019proof}
Peter Ga{\v{z}}i, Aggelos Kiayias, and Dionysis Zindros.
\newblock Proof-of-stake sidechains.
\newblock In {\em 2019 IEEE Symposium on Security and Privacy (SP)}, pages 139--156. IEEE, 2019.

\bibitem{aave}
Aave.
\newblock Aave document hub.
\newblock \url{https://docs.aave.com/hub/}, 2023.

\bibitem{aave-bridge}
Aave.
\newblock Aave governance cross-chain bridges.
\newblock \url{https://github.com/aave/governance-crosschain-bridges}, 2023.

\bibitem{crosschain-loans}
Cross chain Loans.
\newblock Cross-chain loans - decentralized lending marketplace across blockchains.
\newblock \url{https://crosschain.loans/}, 2022.

\bibitem{radar}
DappRadar.
\newblock Staking - radar token - dappradar.
\newblock \url{https://dappradar.com/token/staking}, 2022.

\bibitem{verification}
CoinYuppie.
\newblock In-depth analysis of four types of cross-chain bridges and their risks.
\newblock \url{https://coinyuppie.com/in-depth-analysis-of-four-types-of-cross-chain-bridges-and-their-risks/}, 2022.

\bibitem{wormhole-chain}
Portal~Token Bridge.
\newblock Portal token bridge - introduction.
\newblock \url{https://docs.wormhole.com/wormhole/}, 2023.

\bibitem{orbit-chain}
Orbit Bridge.
\newblock Orbit bridge faqs.
\newblock \url{https://bridge.orbitchain.io/questions}, 2023.

\bibitem{satellite-chain}
Satellite by~Alexar.
\newblock Satellite by alexar docs.
\newblock \url{https://docs.axelar.dev/resources/mainnet}, 2023.

\bibitem{synapse-19}
Synapse.
\newblock Synapse.
\newblock \url{https://synapseprotocol.com/}, 2023.

\bibitem{hop}
Hop.
\newblock Hop exchange.
\newblock \url{https://bridge.connext.network/}, 2023.

\bibitem{across-chain}
Across Protocol.
\newblock Across protocol.
\newblock \url{https://across.to/bridge}, 2023.

\bibitem{boringdao-chain}
BoringDao.
\newblock Boringdao docs.
\newblock \url{https://docs.boringdao.com/}, 2023.

\bibitem{multichain-chain}
Multichain.
\newblock Multichain supported chains.
\newblock \url{https://docs.multichain.org/getting-started/introduction/supported-chains}, 2023.

\bibitem{chainport-chain}
ChainPort.
\newblock Chainport docs.
\newblock \url{https://docs.chainport.io/chainport/chainport-features}, 2023.

\bibitem{thundercore-chain}
ThunderCore.
\newblock Thundercore docs.
\newblock \url{https://docs.developers.thundercore.com/product-protocol/bridges/interact-with-thundercore-bridge}, 2023.

\bibitem{cross-chain}
Cross-Chain Bridge.
\newblock Cross-chain bridge docs - conneted networks.
\newblock \url{https://docs.crosschainbridge.org/connected-networks}, 2023.

\bibitem{hyphen-chain}
Hyphen Bridge.
\newblock Hyphen bridge.
\newblock \url{https://hyphen.biconomy.io/bridge}, 2023.

\bibitem{cbridge-chain}
Celer cBridge.
\newblock Celer cbridge.
\newblock \url{https://cbridge.celer.network/1/56/USDC}, 2023.

\bibitem{nomad-chain}
Nomad.
\newblock Nomad docs.
\newblock \url{https://docs.nomad.xyz/token-bridge/how-to-bridge}, 2023.

\bibitem{allbridge-chain}
Allbridge.
\newblock Allbridge docs.
\newblock \url{https://docs.allbridge.io/allbridge-overview/networks-and-tokens}, 2023.

\bibitem{soy-chain}
Chainspot.
\newblock Soy bridge.
\newblock \url{https://chainspot.io/bridge/soy-bridge}, 2023.

\bibitem{chainlink-ccip}
Chainlink.
\newblock Chainlink: Blockchain oracles for hybrid smart contracts.
\newblock \url{https://chain.link/cross-chain}, 2023.

\bibitem{zarick2021layerzero}
Ryan Zarick, Bryan Pellegrino, and Caleb Banister.
\newblock Layerzero: Trustless omnichain interoperability protocol.
\newblock {\em arXiv preprint arXiv:2110.13871}, 2021.

\bibitem{stargate}
Stargate.
\newblock Welcome to the omnichain future.
\newblock \url{https://stargate.finance}, 2023.

\bibitem{optics}
Optics.
\newblock Bridges by optics v2.
\newblock \url{https://optics.app/}, 2023.

\bibitem{nomad}
Nomad.
\newblock Nomad | bridge.
\newblock \url{https://app.nomad.xyz/}, 2023.

\bibitem{nxtp}
Connext.
\newblock nxtp: A simpler xchain protocol.
\newblock \url{https://blog.connext.network/nxtp-a-simpler-xchain-protocol-88760697ea04}, 2021.

\bibitem{map}
Map Protocol.
\newblock Map protocol.: Secure omnichain layer for web3 \& dapps.
\newblock \url{https://www.mapprotocol.io}, 2023.

\bibitem{0x00000000}
Etherscan.
\newblock Null address.
\newblock \url{ https://etherscan.io/address/0x0000000000000000000000000000000000000000}, 2023.

\bibitem{fusion}
Fusion.
\newblock Distributed control rights management.
\newblock \url{https://www.fusion.org/tech/dcrm}, 2023.

\bibitem{synapse}
Synapse Bridge.
\newblock Welcome to synapse.
\newblock \url{https://docs.synapseprotocol.com}, 2023.

\bibitem{cbridge}
Celer cBridge.
\newblock Welcome to cbridge.
\newblock \url{https://cbridge-docs.celer.network}, 2023.

\bibitem{cbridge-trust}
Celer cBridge.
\newblock State guardian network.
\newblock \url{https://cbridge-docs.celer.network/introduction/state-guardian-network}, 2023.

\bibitem{polygon}
Polygon.
\newblock Polygon - ethereum's internet of blockchains.
\newblock \url{https://polygon.technology/lightpaper-polygon.pdf}, 2023.

\bibitem{near-bridge}
Near.
\newblock Eth <> near rainbow bridge - near protocol.
\newblock \url{https://near.org/bridge/}, 2023.

\bibitem{xdai}
xDAI Bridge.
\newblock xdai bridge - gnosis chain.
\newblock \url{https://bridge.gnosischain.com}, 2023.

\bibitem{polygon-bridge}
Polygon.
\newblock Getting started with polygon proof of stake chain.
\newblock \url{https://wallet.polygon.technology}, 2023.

\bibitem{chainlink}
Chainlink.
\newblock Chainlink: Blockchain oracles for hybrid smart contracts.
\newblock \url{https://chain.link}, 2023.

\bibitem{rugpull-luna}
CoinGeek.
\newblock Solana sees first rug pull: Luna yield disappears with \$6.7m in digital currency.
\newblock \url{https://coingeek.com/solana-sees-first-rug-pull-luna-yield-disappears-with-6-7m-in-digital-currency/}, 2021.

\bibitem{rugpull-blockverse}
United Gamers.
\newblock Rug pull or fud? blockverse disappears, reappears, confuses us all.
\newblock \url{https://www.unitedgamers.gg/news/blockverse-rupull-fud/}, 2022.

\bibitem{wormhole}
Wormhole Bridge.
\newblock Wormhole bridge.
\newblock \url{https://wormholebridge.com/}, 2023.

\bibitem{polynetwork}
PolyNetwork.
\newblock Enhancing connections between ledgers by providing interoperability in web 3.0.
\newblock \url{https://poly.network/}, 2023.

\bibitem{polynetwork-attack}
CertiK.
\newblock Polynetwork attack analysis.
\newblock \url{https://certik.medium.com/polynetwork-hack-analysis-a86513f2a730}, 2021.

\bibitem{plasma}
Plasma Bridge.
\newblock Plasma bridge.
\newblock \url{https://docs.polygon.technology/docs/develop/ethereum-polygon/plasma/getting-started/}, 2023.

\bibitem{plasma-attack}
Coin Culture.
\newblock White hat saves polygon from \$850 million hack.
\newblock \url{https://coinculture.com/au/tech/white-hat-saves-polygon-from-850million-hack/}, 2021.

\bibitem{bnb-attack}
CryptoBriefing.
\newblock Bnb chain’s \$566m hack: Binance network’s major bridge attack unpacked.
\newblock \url{https://cryptobriefing.com/bnb-chain-566m-hack-binance-networks-major-bridge-attack-unpacked/}, 2022.

\bibitem{bnb-hard}
Bitcoin.com.
\newblock Binance-backed blockchain completes hard fork to mitigate future cross-chain bridge hacks.
\newblock \url{https://news.bitcoin.com/binance-backed-blockchain-completes-hard-fork-to-mitigate-future-cross-chain-bridge-hacks/}, 2022.

\bibitem{multichain}
Coin Telegraph.
\newblock Vitalik buterin gives thumbs down to cross-chain applications.
\newblock \url{https://cointelegraph.com/news/vitalik-buterin-gives-thumbs-down-to-cross-chain-applications}, 2022.

\bibitem{multichain-attack}
Halborn.
\newblock Explained: The multichain hack (january 2022).
\newblock \url{https://halborn.com/explained-the-multichain-hack-january-2022/}, 2021.

\bibitem{lifi}
LI.FI.
\newblock Advanced bridge \& dex aggregation.
\newblock \url{https://li.fi}, 2023.

\bibitem{lifi-attack}
LI.FI Blog.
\newblock Li.fi smart contract vulnerability post mortem.
\newblock \url{https://blog.li.fi/20th-march-the-exploit-e9e1c5c03eb9}, 2022.

\bibitem{rubic-attack2}
Binance Feed.
\newblock Rubic lost more than \$1.4 million due to the hack.
\newblock \url{https://www.binance.com/en/feed/post/134659}, 2022.

\bibitem{chainswap}
ChainSwap.
\newblock The cross-chain hub for all ecosystems.
\newblock \url{https://chainswap.com}, 2023.

\bibitem{razor}
RazorNetwork.
\newblock Truly decentralized oracle network for decentralized.
\newblock \url{https://razor.network}, 2023.

\bibitem{wormhole-attack}
Chainalysis.
\newblock Lessons from the wormhole exploit: Smart contract vulnerabilities introduce risk; blockchains’ transparency makes it hard for bad actors to cash out.
\newblock \url{https://blog.chainalysis.com/reports/wormhole-hack-february-2022/}, 2022.

\bibitem{multichain-attack2}
REKT Database.
\newblock Multichain attack.
\newblock \url{https://de.fi/rekt-database/multichain}, 2023.

\bibitem{wonderhero-attack}
Metaverse Post.
\newblock Wonderhero token collapses after hack.
\newblock \url{https://mpost.io/wonderhero-token-collapses-after-hack/}, 2022.

\bibitem{ronin}
Ronin Network.
\newblock Ronin network.
\newblock \url{https://bridge.roninchain.com/}, 2023.

\bibitem{ronin-attack}
HALBORN.
\newblock Explained: The ronin hack (march 2022).
\newblock \url{https://halborn.com/explained-the-ronin-hack-march-2022/}, 2022.

\bibitem{sky-mavis}
Sky Mavis.
\newblock Blockchain: Gamified.
\newblock \url{https://www.skymavis.com/}, 2023.

\bibitem{axie-dao}
Axie DAO.
\newblock We are the axie dao.
\newblock \url{https://axiedao.org/}, 2023.

\bibitem{harmony}
Horizon by~Harmony.
\newblock Harmony one-eth bridge.
\newblock \url{https://bridge.harmony.one/busd}, 2023.

\bibitem{harmony-attack}
HALBORN.
\newblock Harmony's horizon bridge attack: How \$100m was siphoned off by a hacker.
\newblock \url{https://hackernoon.com/harmonys-horizon-bridge-attack-how-dollar100m-was-siphoned-by-a-hacker}, 2022.

\bibitem{qan-attack}
The Block.
\newblock Hacker drains \$2 million from qan platform bridge, token slumps 94\%.
\newblock \url{https://www.theblock.co/post/176118/hacker-drains-1-million-from-qanplatform-bridge-token-slumps-94}, 2022.

\bibitem{rubic-attack}
Cryptonomist.
\newblock Rubic dex loses \$1 million in crypto to hacker attack.
\newblock \url{https://en.cryptonomist.ch/2022/11/03/dex-rubic-loses-1-million-crypto/}, 2022.

\bibitem{pnetwork-attack2}
Crypto News.
\newblock pnetwork clears the air after rumored over \$1b breach on their platform.
\newblock \url{https://crypto.news/pnetwork-clears-the-air-after-rumored-over-1b-breach-on-their-platform/}, 2022.

\bibitem{polynetwork-attack2}
Rekt Database.
\newblock Bridge attacks.
\newblock \url{https://de.fi/rekt-database}, 2023.

\bibitem{polkabridge}
PolkaBridge.
\newblock First cross-chain \& multichain amm.
\newblock \url{https://polkabridge.org}, 2023.

\bibitem{poolz-finance-attack}
Numen~Cyber Labs.
\newblock Poolz finance attacked for \$665,000.
\newblock \url{https://medium.com/@numencyberlabs/poolz-finance-attacked-for-665-000-56084cacae53}, 2023.

\bibitem{nomad-attack-certik}
CertiK.
\newblock Nomad bridge exploit incident analysis.
\newblock \url{https://www.certik.com/resources/blog/28fMavD63CpZJOKOjb9DX3-nomad-bridge-exploit-incident-analysis}, 2022.

\bibitem{0xa8c83b1b}
Etherscan.
\newblock Address.
\newblock \url{ https://etherscan.io/address/0xa8c83b1b30291a3a1a118058b5445cc83041cd9d}, 2023.

\bibitem{0xcca9299c}
Moonscan.
\newblock Moonbeam transaction hash (txhash) details.
\newblock \url{ https://moonscan.io/tx/0xcca9299c739a1b538150af007a34aba516b6dade1965e80198be021e3166fe4c}, 2022.

\bibitem{0xa5fe9d04}
Etherscan.
\newblock Ethereum transaction hash (txhash) details.
\newblock \url{ https://etherscan.io/tx/0xa5fe9d044e4f3e5aa5bc4c0709333cd2190cba0f4e7f16bcf73f49f83e4a5460}, 2022.

\bibitem{usdt}
Coinbase Help.
\newblock Tether (usdt) - coinbase help.
\newblock \url{https://help.coinbase.com/en/coinbase/getting-started/crypto-education/usdt}, 2023.

\bibitem{qubit}
Qubit.
\newblock Driving growth with personalization. make ecommerce more personal.
\newblock \url{https://www.qubit.com}, 2023.

\bibitem{meter}
Meter.
\newblock The future is multi-chain.
\newblock \url{https://meter.io}, 2023.

\bibitem{meter-attack}
Meter.io.
\newblock Community, unfortunately meter passport was hacked a few hours ago.
\newblock \url{https://twitter.com/Meter_IO/status/1490045486606139392}, 2022.

\bibitem{omni-attack}
Neptune Mutual.
\newblock Decoding omni bridge’s call data replay exploit.
\newblock \url{https://medium.com/neptune-mutual/decoding-omni-bridges-call-data-replay-exploit-f1c7e339a7e8}, 2022.

\bibitem{THORChain-attack}
SlowMist.
\newblock Slowmist: Analysis of three consecutive attacks on thorchain (released in 2021).
\newblock \url{https://slowmist.medium.com/slowmist-analysis-of-three-consecutive-attacks-on-thorchain-6223f1c691be}, 2021.

\bibitem{pnetwork-attack}
Halborn.
\newblock Explained: The pnetwork hack (september 2021).
\newblock \url{https://halborn.com/explained-the-pnetwork-hack-september-2021/}, 2021.

\bibitem{cennznet}
Cennznet.
\newblock Blockchain for the open metaverse.
\newblock \url{https://cennz.net}, 2023.

\bibitem{evodefi-attack}
BanklessTimes.
\newblock Crypto platform evodefi loses \$320k in hack.
\newblock \url{https://www.banklesstimes.com/news/2022/03/22/crypto-platform-evodefi-loses-dollar320k-in-hack/}, 2022.

\bibitem{celer-attack}
SlowMist.
\newblock Truth behind the celer network cbridge cross-chain bridge incident: Bgp hijacking.
\newblock \url{https://medium.com/coinmonks/truth-behind-the-celer-network-cbridge-cross-chain-bridge-incident-bgp-hijacking-52556227e940}, 2022.

\bibitem{wormhole-after}
DEFIYIELD.App.
\newblock Wormhole exploit: the second-largest defi hack ever.
\newblock \url{https://blog.defiyield.app/wormhole-exploit-the-second-largest-defi-hack-ever-237ed5c81670}, 2022.

\bibitem{multichain-after}
Multichain.
\newblock Multichain contract vulnerability post mortem.
\newblock \url{https://medium.com/multichainorg/multichain-contract-vulnerability-post-mortem-d37bfab237c8}, 2022.

\bibitem{multichain-after2}
Coin Telegraph.
\newblock Multichain asks users to revoke approvals amid ‘critical vulnerability’.
\newblock \url{https://cointelegraph.com/news/multichain-asks-users-to-revoke-approvals-amid-critical-vulnerability}, 2022.

\bibitem{ronin-after}
Crypto Potato.
\newblock Ronin network announces bridge restart date three months after \$625m hack.
\newblock \url{https://cryptopotato.com/ronin-network-announces-bridge-restart-date-three-months-after-625m-hack/}, 2022.

\bibitem{mossberg2019manticore}
Mark Mossberg, Felipe Manzano, Eric Hennenfent, Alex Groce, Gustavo Grieco, Josselin Feist, Trent Brunson, and Artem Dinaburg.
\newblock Manticore: A user-friendly symbolic execution framework for binaries and smart contracts.
\newblock In {\em 2019 34th IEEE/ACM International Conference on Automated Software Engineering (ASE)}, pages 1186--1189. IEEE, 2019.

\bibitem{horizon-bridge}
Harmony.
\newblock Horizon bridge - harmony.
\newblock \url{https://docs.harmony.one/home/general/bridges}, 2023.

\bibitem{hamster}
CoinCodeCap.
\newblock Hamster coin rugged: Reportedly 1,730 bnb lost.
\newblock \url{https://coincodecap.com/hamster-coin-rugged-reportedly-1730-bnb-lost}, 2022.

\bibitem{tornado}
Coindesk.
\newblock Us treasury's tornado cash sanctions are ‘unprecedented,’ warns congressman.
\newblock \url{https://www.coindesk.com/policy/2022/08/18/treasury-tornado-cash-sanctions-are-unprecedented-warns-us-congressman/}, 2022.

\bibitem{sereum-ndss19}
Michael Rodler, Wenting Li, Ghassan Karame, and Lucas Davi.
\newblock Sereum: Protecting existing smart contracts against re-entrancy attacks.
\newblock In {\em Proceedings of the 26th Network and Distributed System Security Symposium}, 2019.

\bibitem{grossman2017online}
Shelly Grossman, Ittai Abraham, Guy Golan-Gueta, Yan Michalevsky, Noam Rinetzky, Mooly Sagiv, and Yoni Zohar.
\newblock Online detection of effectively callback free objects with applications to smart contracts.
\newblock {\em Proceedings of the ACM on Programming Languages}, 2017.

\bibitem{frank2020ethbmc}
Joel Frank, Cornelius Aschermann, and Thorsten Holz.
\newblock $\{$ETHBMC$\}$: A bounded model checker for smart contracts.
\newblock In {\em 29th USENIX Security Symposium (USENIX Security 20)}, pages 2757--2774, 2020.

\bibitem{torres2018osiris}
Christof~Ferreira Torres, Julian Sch{\"u}tte, and Radu State.
\newblock Osiris: Hunting for integer bugs in ethereum smart contracts.
\newblock In {\em Proceedings of the 34th Annual Computer Security Applications Conference}, pages 664--676, 2018.

\bibitem{Zhang2020TXSPECTORUA}
Mengya Zhang, Xiaokuan Zhang, Yinqian Zhang, and Zhiqiang Lin.
\newblock Txspector: Uncovering attacks in ethereum from transactions.
\newblock In {\em USENIX Security Symposium}, 2020.

\bibitem{jiang2018contractfuzzer}
Bo~Jiang, Ye~Liu, and WK~Chan.
\newblock Contractfuzzer: Fuzzing smart contracts for vulnerability detection.
\newblock In {\em 2018 33rd IEEE/ACM International Conference on Automated Software Engineering (ASE)}, pages 259--269. IEEE, 2018.

\bibitem{grieco2020echidna}
Gustavo Grieco, Will Song, Artur Cygan, Josselin Feist, and Alex Groce.
\newblock Echidna: effective, usable, and fast fuzzing for smart contracts.
\newblock In {\em Proceedings of the 29th ACM SIGSOFT International Symposium on Software Testing and Analysis}, 2020.

\end{thebibliography}
}

\newpage
{
\appendix

\subsection{Cross-chain Bridge Attack Surfaces}
\label{app:surf}

\bheading{A7: Vulnerable bridge smart contracts.}
Common vulnerabilities in smart contracts (e.g., access control) are possible in bridge smart contracts. We illustrate some potential attacks as follows, taking the liquidity pool smart contract code as an example in~\figref{figs:lpsc}. 

\begin{figure}[!htbp]
\setlength{\fboxsep}{0pt}%
\setlength{\fboxrule}{0pt}%
\tiny
\centering
\begin{lstlisting}[language=Solidity]
contract LiquidityPool {
  address bridgeAdmin;
  mapping(address => uint256) balances;
  uint256 liquidReserves;
  
  modifier onlyAdmin {
    require(msg.sender == bridgeAdmin);
  }

  constructor(address _bridgeAdmin) {
    bridgeAdmin = _bridgeAdmin;
  }
  
  function changeAdmin(address _newAdmin) public onlyAdmin {
    bridgeAdmin = _newAdmin;
  }

  function addLiquidity(uint256 tokenAmount) public payable {
    require(balances[msg.sender] >= tokenAmount);
    uint256 lpTokenAmount = tokenAmount / exchangeRatio();
    mint(msg.sender, lpTokenAmount);
    liquidReserves += tokenAmount;
    emit AddLiquidity(tokenAmount, lpTokenAmount,   msg.sender);
  }

  function removeLiquidity(uint256 lpTokenAmount) public {
    uint256 tokenAmountReturn = lpTokenAmount * exchangeRatio();
    require(liquidReserves >= tokenAmountReturn);
    burn(msg.sender, lpTokenAmount);
    liquidReserves -= tokenAmountReturn;
    emit RemoveLiquidity(tokenAmountReturn, lpTokenAmount, msg.sender);
  }
}
\end{lstlisting}
\caption{An example of liquidity pool smart contract.}
\label{figs:lpsc}
\end{figure}
\begin{packeditemize}
    \item Incorrect initialization. Bridge accounts will be set as the admin/owner addresses of the liquidity pool during the pool contract creation. If these privileged addresses are initialized with incorrect values in Line 10-11, the total assets in the smart contract can be drained. 
    \item Inappropriate function permission. Similarly, there are some special functions that change privileged addresses or control contracts, such as killing or pausing contracts when the caller is an admin/owner. If these functions are created with inappropriate permissions in Line 14-15 (e.g., allowing anyone to call them), the contracts will be hacked.
    \item Unchecked balance. To add liquidity to the pools, the liquidity providers should transfer their tokens. If the balance of tokens is not checked in Line 19, attackers can deposit nothing or tokens with less value than the value of returned tokens (e.g., LP tokens). Similarly, there will be problems if balance check is not performed correctly in Line 28.
    \item Miscalculated token price. If the amount of LP tokens is given to the liquidity providers more than desired due to the miscalculated token price in Line 20, the pool will lose money. In particular, the function \textit{exchangeRatio} calculates the token price. Similarly, in the liquidity remove process, if the amount of original tokens returned to the liquidity providers is more than desired in Line 27, there will be financial losses too. 
    \item Inconsistent event. The events will be emitted for off-chain bridge server to monitor, when non-trivial actions happen. For example, \textit{AddLiquidity} event is emitted in Line 23 when the function \textit{addLiquidity} is called. If the \textit{tokenAmount} representing the amount of deposited tokens is wrong, the bridge will get incorrect information and there will be inconsistencies. Similarly, the event \textit{RemoveLiquidity} in Line 31 should be correctly recorded. 
\end{packeditemize}

}
\end{document}